\documentclass[aps,prl,twocolumn,superscriptaddress]{revtex4-1}

\usepackage{amsmath}
\usepackage{amssymb}
\usepackage{mathtools}
\usepackage{graphicx}
\usepackage[usenames,dvipsnames]{color}
\usepackage{bm}
\usepackage{hyperref}
\usepackage{url}
\usepackage[utf8]{inputenc}
\usepackage{subfigure}
\usepackage{slashed,bbm}
\usepackage{graphics,psfrag,epsfig}
\usepackage{dsfont}
\usepackage{setspace}
\usepackage{wasysym}

\usepackage{hyperref}
\usepackage{xcolor}
\definecolor{dark-red}{rgb}{0.4,0.15,0.15}
\definecolor{dark-blue}{rgb}{0.15,0.15,0.4}
\definecolor{medium-blue}{rgb}{0,0,0.5}
\hypersetup{
colorlinks, linkcolor={dark-blue},
citecolor={dark-blue}, urlcolor={medium-blue}
}
\newcommand{\be}{\begin{equation}}
\newcommand{\ee}{\end{equation}}

\newcommand{\XY}{\text{XY}}

\newcommand{\vF}{\upsilon_\text{F}}
\newcommand{\vD}{\upsilon_\Delta}
\newcommand{\vs}{\upsilon_\text{s}}
\newcommand{\Eg}{E_\text{g}}
\newcommand{\Es}{E_\text{s}}
\newcommand{\dk}{\text{d}k}
\renewcommand{\i}{\text{i}}

\usepackage[normalem]{ulem}
\usepackage{slashed,bbm}

\begin{document}

\title{Competing nodal d-wave superconductivity and antiferromagnetism}

\author{Xiao Yan Xu}
\affiliation{Department of Physics, University of California at San Diego, La Jolla, California 92093, USA}
\author{Tarun Grover}
\affiliation{Department of Physics, University of California at San Diego, La Jolla, California 92093, USA}

\date{\today}

\begin{abstract}
	Competing unconventional superconductivity and antiferromagnetism widely exist in several strongly correlated quantum materials whose direct simulation generally suffers from fermion sign problem. Here we report unbiased Quantum Monte Carlo (QMC) simulations  on a sign-problem-free repulsive 	toy model with same onsite symmetries as the standard Hubbard model on a 2D square lattice. Using QMC, supplemented with mean-field and continuum field-theory arguments, we find that it hosts three distinct phases: a nodal d-wave phase, an antiferromagnet, and an intervening phase which hosts coexisting antiferromagnetism and nodeless d-wave superconductivity. The transition from the coexisting phase to the antiferromagnet is described by the 2+1-D XY universality class, while the one from the coexisting phase to the nodal d-wave phase is described by the  Heisenberg-Gross-Neveu theory. The topology of our phase diagram resembles that of layered organic materials which host pressure tuned Mott transition from antiferromagnet to unconventional superconductor at half-filling.  
\end{abstract}

\maketitle
{\it Introduction}\,---\,
The interplay between unconventional superconductivity and magnetism plays a crucial role in a wide variety of strongly correlated systems~\cite{Norman2011} such as 
cuprates~\cite{Dagotto94, Lee06_rev}, heavy fermions \cite{steglich1984heavy,Stewart1984,
Amato1997,Pfleiderer2009,Pagliuso2001,maple_rev2010,Steglich79,Joynt2002,Knebel2004,
Park2006,Yuan2003,Mito2003,Aeppli1987}, layered organic conductors~\cite{Kagawa05,Kurosaki2005,Lefebvre00,Arai2001,
Belin1998,DeSoto1995,Mayaffre1995,
Kanoda1996,Jeerome1991}, iron-based superconductors \cite{stewart2011superconductivity, si2016high, Dai2015, Pratt2009, Nandi2010}, Helium-3 \cite{vollhardt2013superfluid} and even in recently studied twisted 2D materials \cite{balents2020superconductivity}. Superconductors with nodal quasiparticles are particularly interesting since the fermionic quasiparticles cannot be neglected even for the ground state properties of the superconductor, or for understanding quantum phase transitions to proximate phases. 

One route to obtain an unconventional superconductor is to dope a Mott insulator~\cite{Anderson87,Lee06_rev}, as is experimentally the case for nodal d-wave superconductor (dSC) in cuprates. From a numerical perspective, this is rather challenging: the combination of `Mottness' and continuously varying filling leads to the fermion sign problem, making the unbiased simulation of competing nodal superconductivity and antiferromagnetism an unrealized goal. 
However, one notes that two of the most prominent symmetry breaking phases in cuprates, namely the dSC or the AFM insulator in principle do \textit{not} require doping for their existence. Therefore, the unbiased large-scale Quantum Monte Carlo (QMC) simulations of quantum transitions between dSC and AFM is possible.
From an experimental perspective, competing dSC and AFM phases at half-filling are relevant to organic materials ~\cite{Kagawa05,Kurosaki2005,Lefebvre00,Arai2001,
	Belin1998,DeSoto1995,Mayaffre1995,
	Kanoda1996,Jeerome1991}, such as the layered organic material $\kappa$-$\text{ET}_2\text{Cu[N(CN)}_2]\text{Cl}$, which undergoes a transition from an AFM Mott insulator to a superconductor under pressure.


\begin{figure}[t]
  \centering
	\includegraphics[width=0.95\hsize]{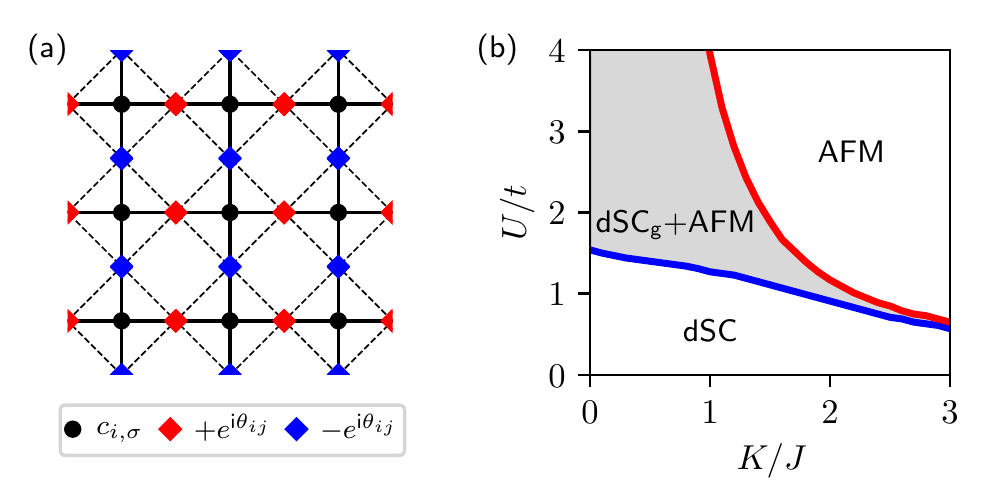}
	\caption{The lattice model and mean-field phase diagram at $V/t=0.5$. (a) The lattice model. Fermions (black circle dots) live on a square lattice with nearest hopping and onsite Hubbard interaction $U$.  Rotors (colored square dots) live on the bonds of the same square lattice described by a quantum rotor Hamiltonian. Rotors coupled to fermion pairing operators with the d-wave form indicated by different signs ($\tau_{ij}=\pm 1$) in front of $e^{\i\theta_{ij}}$.    (b) Mean-field phase diagram. dSC stands for the nodal $d_{x^2-y^2}$ superconductor, AFM for the antiferromagnetic insulator,  and dSC$_\text{g}$+AFM for the coexisting gapped d-wave superconductor and antiferromagnetic insulator. }
	\label{fig:fig1}
\end{figure}

In this Letter, we will propose a model which does not suffer from the fermion sign problem, and which demonstrably hosts both a nodal dSC, and an AFM insulator on a 2D square lattice, with an intermediate phase with coexisting of AFM and nodeless dSC (denoted as dSC$_\text{g}$+AFM), through state-of-the-art Determinantal QMC (DQMC) simulations.
 The phase transition between dSC$_\text{g}$+AFM and AFM appears to be continuous and in the 3D XY universality class. The phase transition
between nodal dSC and dSC$_\text{g}$+AFM also appears to be continuous, which we will argue below to lie in the Heisenberg-Gross-Neveu (HGN) universality class.

{\it Model}\,---\,
The Hilbert space of our model consists of spinful fermions $c_{i,\sigma}$ that live on the vertices $\{i\}$ of a square lattice, and fluctuating cooper pairs $e^{\i\theta_{ij}}$ (bosons) living on the bonds $\{ij\}$ of the same square lattice, as shown in Fig.~\ref{fig:fig1}(a).
The Hamiltonian is given by:

\be 
H = H_t + H_U + H_V +  H_\XY 
\ee 
 Here $H_t + H_U = -t\sum_{\langle ij\rangle,\sigma}(c_{i,\sigma}^{\dagger}c_{j,\sigma}+\text{h.c.}) + \frac{U}{2}\sum_{i}(\rho_{i,\uparrow}+\rho_{i,\downarrow}-1)^2$ is the standard Hubbard model with nearest neighbor hopping, with $\rho_{i,\sigma} = c_{i,\sigma}^{\dagger} c_{i,\sigma}$. $H_V$ is the coupling between fermion operator corresponding to  d-wave pairing and rotors $\theta$,
$
H_{V}=V\sum_{\langle ij\rangle}(\tau_{i,j}e^{\i\theta_{ij}}(c_{i,\uparrow}^{\dagger}c_{j,\downarrow}^{\dagger}-c_{i,\downarrow}^{\dagger}c_{j,\uparrow}^{\dagger})+\text{h.c.})$
with $\tau_{i,i\pm\hat{x}}=1$, $\tau_{i,i\pm\hat{y}}=-1$. Finally, $H_\XY$ is a quantum rotor Hamiltonian describing the dynamics and self-interactions of 
cooper pairs:
$
H_{\XY}=K\sum_{\langle ij\rangle}n_{ij}^{2}-J\sum_{\langle ij,il\rangle}\cos(\theta_{ij}-\theta_{il}),
$
where $n_{ij}$ is the angular momentum of rotors, $[n_{ij},e^{\pm \i \theta_{ij}}]=\pm e^{\pm \i \theta_{ij}} $, and $\sum_{\langle ij,il\rangle}$ denotes summation over all pairs of nearest neighbor bonds, i.e., bonds that share a site. We note the $K$-term is the kinetic part of the fluctuating cooper pairs, and the $J$-term will be generated dynamically.
The model has the same symmetries as the conventional Hubbard model: a spin-rotation SU(2) symmetry and a charge U(1) symmetry given by $c_{i,\sigma} \rightarrow c_{i,\sigma} e^{\i\varphi} $, $\theta \rightarrow \theta + 2\varphi$. It also has a particle-hole symmetry $c_{i,\sigma} \rightarrow \epsilon_i c_{i,\sigma}^{\dagger}$, $\theta \rightarrow -\theta$, at half-filling, where $\epsilon_i=(-)^i$.
Perhaps most importantly, the model also hosts an  anti-unitary symmetry $\mathcal{U}:\ c_{i,\uparrow} \rightarrow \epsilon_i c_{i,\downarrow}^\dagger,\ c_{i,\downarrow} \rightarrow -\epsilon_i c_{i,\uparrow}^\dagger, \ \i \rightarrow -\i$ and $\mathcal{U}^2=-1$, which makes the model sign problem-free ~\cite{wuzhang2005}.

One may obtain some features of the global phase diagram of our model without detailed calculations. Setting $U = 0$, when $K/J \ll 1$, the charge-$U(1)$ symmetry will be spontaneously broken, and therefore, the fermion-part of the Hamiltonian reduces to the BCS mean-field theory for a nodal $d_{x^2-y^2}$ superconductor, which we denote as dSC. The dSC phase is expected to be stable at small $U$ since weak interactions are irrelevant for the nodal Dirac fermions. When $K/J$ increases, eventually the charge-$U(1)$ symmetry is expected to get restored due to fluctuations of the $\theta$ field. Since the unit-cell of $H$ contains an odd number of fermions, a gapped trivial paramagnet is ruled out \cite{Oshikawa00a, Hasting04}, and energetically, we expect that the phase at $U/t \gg 1$ and $K/J \gg 1$ to be a conventional antiferromagnet. Furthermore, since the nodes in dSC are separated by $(\pi, \pi)$, which is the ordering wavevector for AFM, one may also expect a phase where AFM coexists with a \textit{gapped} d-wave superconductor. As well we see, these expectations are born out by the DQMC calculations, but first, we consider a mean-field theory.

\begin{figure}[t]
  \centering
	\includegraphics[width=0.95\hsize]{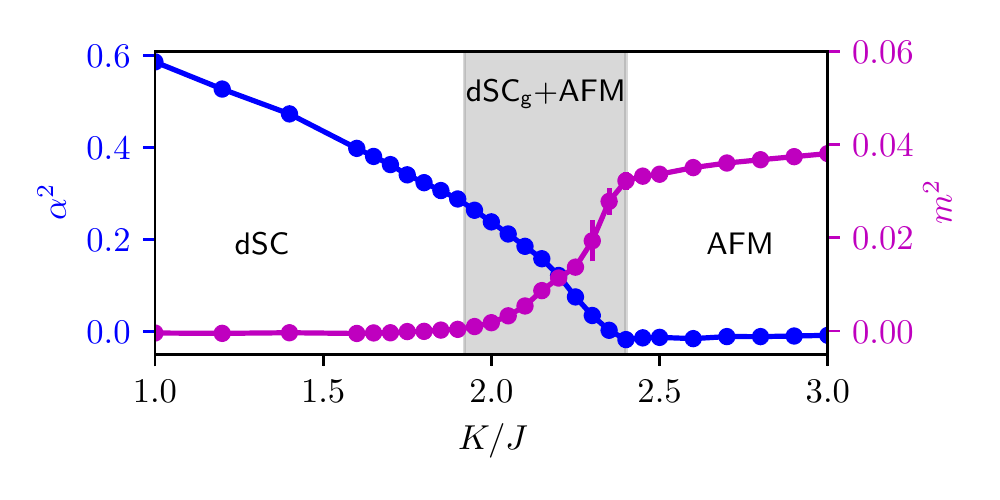}
	\caption{QMC Phase diagram of our model at $V/t=0.5$ and $U/t=4.0$. Here $\alpha$ is the dSC order parameter, $m$ is the AFM order parameter. }
	\label{fig:fig2}
\end{figure}

\begin{figure*}[t]
  \centering
	\includegraphics[width=1.0\hsize]{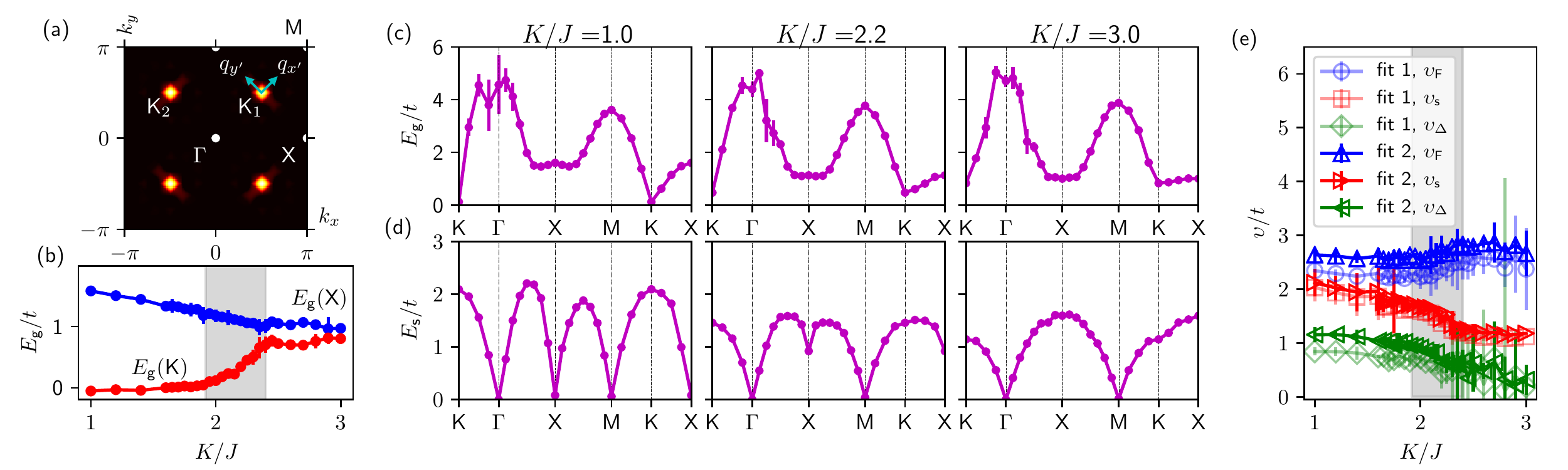}
	\caption{Single-particle and spin gaps and velocities. (a) The first Brillouin zone (BZ) of the square lattice. Four light points are the location of nodes, and the lightness of the color denotes integrated spectral weight $A(\vec{k},\omega)$ over a small energy window near the Fermi level $(0,0.5t)$. (b) Single-particle gap at the nodal point (K) and the anti-nodal point (X) extrapolated to thermodynamic limit. (c) Single-particle gap along the path K-$\Gamma$-X-M-K-X of BZ at $L=16$. (d) Spin gap along the same path of BZ at $L=16$. (e) Fermi velocity, spin-wave velocity and pairing velocity extracted using two different fitting schemes.}
	\label{fig:fig3}
\end{figure*}

{\it Mean-field phase diagram}\,---\,
Defining $\hat{\alpha}_{ij}\equiv -e^{i\theta_{ij}}$ and
$\hat{\Delta}_{ij}\equiv \tau_{i,j} ( c_{i,\uparrow}^\dagger c_{j,\downarrow}^\dagger - c_{i,\downarrow}^\dagger c_{j,\uparrow}^\dagger ) $, we are led to two coupled mean-field Hamiltonians, one for the fermions, $H_\text{f}^{\text{MF}}$ and the other for the rotors, $H_{\theta}^{\text{MF}}$: $H_\text{f}^{\text{MF}}  =  -t\sum_{\langle ij \rangle,\sigma}\left(c_{i,\sigma}^{\dagger}c_{j,\sigma}+h.c.\right)-Um\sum_{i}(-)^{i}(\rho_{i,\uparrow}-\rho_{i,\downarrow})
- V\alpha\sum_{\langle ij \rangle} \left( \hat{\Delta}_{ij} + \text{h.c.} \right)$, and $H_{\theta}^{\text{MF}} =  K\sum_{\langle ij\rangle}n_{ij}^{2}-J\sum_{\langle ij,il\rangle}\cos(\theta_{ij}-\theta_{il})
- V\Delta\sum_{\langle ij \rangle} \left( \hat{\alpha}_{ij}+\text{h.c.} \right)$.
Here $\alpha = \langle \hat{\alpha}_{ij} \rangle $, $\Delta = \langle \hat{\Delta}_{ij} \rangle$ and $m = \langle (-)^{i}(\rho_{i,\uparrow}-\rho_{i,\downarrow}) \rangle$, and we have chosen the antiferromagnetic order parameter to point along the $z$-axis in the spin-space. We can solve the two coupled mean-field Hamiltonian self-consistently. 
The $H_{\theta}^{\text{MF}}$ part is still an interacting rotor lattice problem which we solve using numerical exact diagonalization (ED) on a small cluster consisting of four bonds of a square lattice.
After we obtain the value of $\alpha$, we solve the fermion part $H_{\text{f}}^{\text{MF}}$, and find a self-consistent solution. We set $t=1$, $J=1$, $V/t=0.5$, and explore the $U/t-K/J$ phase diagram as shown in Fig.~\ref{fig:fig1}(b). At small $U/t$, the (nodal) dSC is a stable phase, while at larger $U/t$, we enter a phase with coexisting gapped d-wave superconductivity (dSC$_\text{g}$) and antiferromagnetism (AFM). At still larger $U/t$, a pure AFM phase without any superconductivity is stabilized. Tuning $K/J$  changes the relative sizes of these three phases. Overall, when we increase $K/J$, the pure AFM region becomes larger, while the coexistence phase region (dSC$_\text{g}$+AFM) shrinks.
We note that a variational cluster perturbation theory and cluster dynamical mean-field theory calculations on doped repulsive Hubbard model also find a coexistence phase similar to ours (dSC$_\text{g}$+AFM)~\cite{Lichtenstein2000,Jarrell2001,Senechal2005,Capone2006,Nevidomskyy2008,Kancharla2008}.

{\it QMC phase diagram}\,---\,
The model can be simulated with the DQMC method without sign problem (See Refs.~\cite{suppl,BSS,AssaadEvertz2008} for additional technical details of DQMC). In DQMC, the imaginary time evolution is Trotter decomposed into $L_\tau$ slices, $\beta t =L_\tau \Delta_\tau$, where imaginary time step $\Delta_\tau=0.1$ is used in our simulations. We employ the standard Hubbard-Stratonovich (HS) transformation to decouple the repulsive Hubbard interaction into fermion bilinears coupled to auxiliary fields \cite{Hirsch83}. To explore the ground state properties,  we scale the inverse temperature with the linear system size $L$, in particular, we fix $\beta t = 2L$ and perform simulations up to $L=20$.  Motivated by the mean-field phase diagram in the $U/t$-$K/J$ plane, one only needs to tune one parameter to explore all  three possible phases. We fix parameters $U/t=4.0$, $V/t=0.5$, and explore possible phases by tuning only $K/J$. 
The AFM order parameter, $\vec{m}=\langle (-)^{i} \vec{S}_i \rangle$, and the dSC order parameter, 
$\alpha = \langle \hat{\alpha}_{ij}\rangle$, are extracted from static correlation functions $m^2 = \frac{1}{L^4} \sum_{i,j} (-)^{i+j} \left\langle\vec{S}_i \cdot \vec{S}_j \right\rangle$ and  $\alpha^2 = \frac{1}{4L^4} \sum_{\langle ij \rangle, \langle k l \rangle} \langle e^{\i \theta_{ij}} e^{\i \theta_{kl}} \rangle$. Fig.~\ref{fig:fig2} shows these order parameters extrapolated to the thermodynamic limit, see \cite{suppl} for technical details. For $K/J<1.92(5)$, only $\alpha \ne 0$, which corresponds to the nodal dSC phase. This is also evident from the spectral function integrated over a small energy window, Fig.\ref{fig:fig3}(a), which shows four distinct nodes.
For $1.92(5)<K/J<2.40(5)$, we have both $\alpha \ne 0$ and $m \ne 0$, and therefore this is the dSC$_\text{g}$+AFM phase. For $K/J>2.40(5)$, we enter the pure AFM phase where only $m\ne0$. 

The phase diagram is also consistent with the results for the single-particle gap as well as the spin-gap, see Figs. \ref{fig:fig3}(b), (c), (d). As shown in Fig.~\ref{fig:fig3}(c), in the nodal dSC region $K/J<1.92(5)$, we have nodes at the K points, while the antinodal points X are gapped. The single particle gaps open both in the dSC$_\text{g}$+AFM and the AFM regions. Similarly, the gap to spinful-excitations remains zero at $\Gamma$ and M points in the  dSC$_\text{g}$+AFM and the AFM regions, due to the Goldstone modes resulting from the spin-rotation symmetry breaking.

\begin{figure}[t]
  \centering
	\includegraphics[width=0.95\hsize]{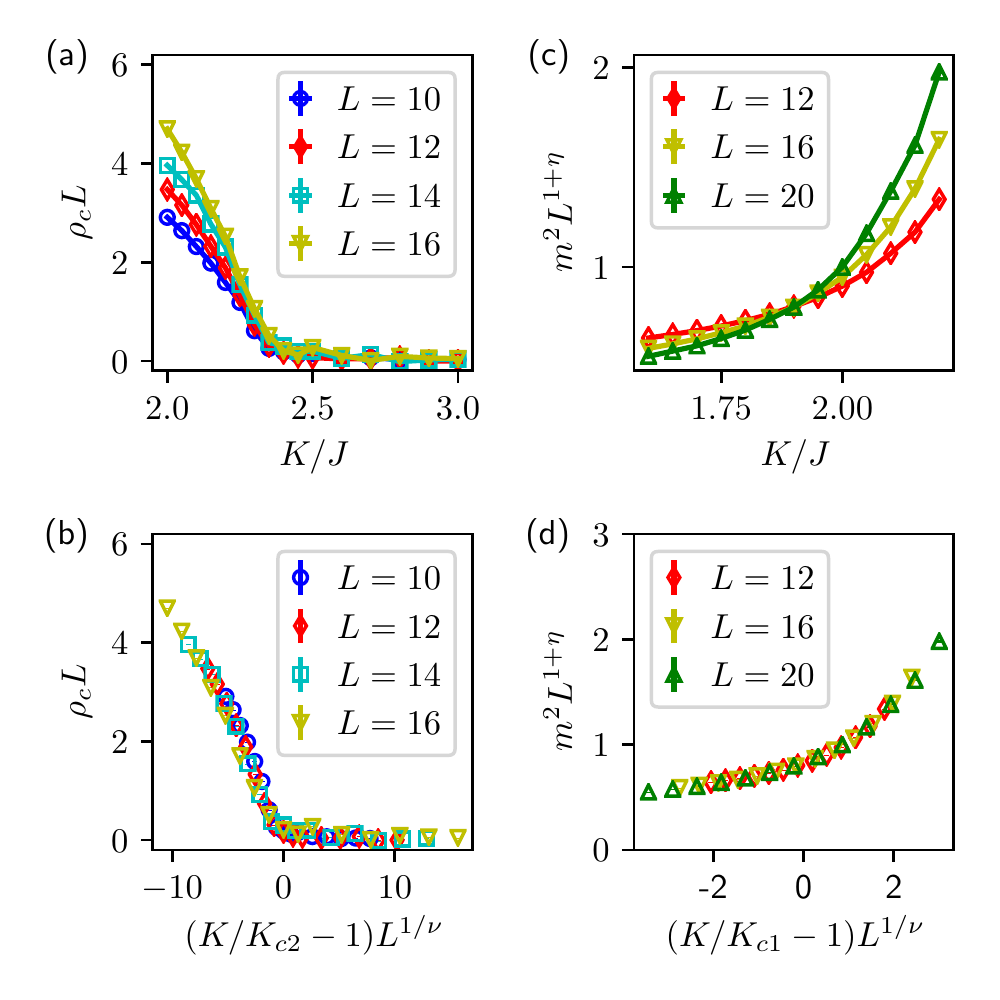}
	\caption{Data collapse to obtain critical exponents. (a) Charge stiffness $\rho_c$ for different system sizes near the second phase transition from the coexistence phase to the AFM phase. (b) Data collapse of the charge stiffness in (a) with $K_{c2}=2.40(5)$ leads to  $\nu\approx0.67$. (c) Squared AFM order parameter $m^2$ near the transition from the nodal dSC phase to the coexistence phase. (d) Data collapse of $m^2$ in (c) with $K_{c1}=1.92(5)$ leads to $\nu=0.99(8)$ and $\eta=0.55(2)$.}
	\label{fig:fig4}
\end{figure}

{\it Phase transitions}\,---\,
The transitions from the dSC to the dSC$_\text{g}$+AFM and from the dSC$_\text{g}$+AFM to the AFM both appear to be continuous.
The transition from the dSC$_\text{g}$+AFM to the AFM is the conventional XY transition, and the data for the charge stiffness can be collapsed quite well with 3D XY exponents, see Fig.~\ref{fig:fig4}(a) and (b).
The transition from the dSC to the dSC$_\text{g}$+AFM is a more interesting one, which on theoretical ground we believe to be in the  HGN class (see below). Using data collapse, we extract the critical exponents for this transition, as shown in Figs.~\ref{fig:fig4}(c) and (d). We estimate  the correlation length exponent $\nu=0.99(8)$ and the anomalous dimension of the AFM order-parameter $\eta_{m}=0.55(2)$.

We also measured the Fermi velocity $\upsilon_F$, the dSC pairing velocity $\upsilon_\Delta$ defined via the nodal dispersion $\Eg(\vec{q}')=\sqrt{\upsilon_F^2q_x'^2+\upsilon_\Delta^2q_y'^2+\Eg^2(\text{K})}$, and the spin-wave velocity $\vs$ defined via the dispersion of Goldstone modes at the $\text{M}$ point through $\Es(\vec{q}')=\sqrt{\upsilon_s^2 q'^2 + \Es^2(\text{M})}$. To extract these velocities, we considered two different finite size scaling schemes. In scheme 1, we fix $\delta q = \frac{2\pi}{L_\text{max}}$, and for system sizes less than $L_\text{max}=20$, $\Eg(\text{K}+\vec{\dk}_1)$, $\Eg(\text{K}+\vec{\dk}_2)$ and $\Es(\Gamma+\vec{\dk}_1)$ are obtained by interpolation. After we obtain gap functions for each system size, we perform an $1/L$ extrapolation of the gap, and finally use the above formulas to obtain the velocities.
In scheme 2, we first calculate the velocities based on the above formulas, and then perform the $1/L$ extrapolation of the velocities. The velocities obtained from these schemes are shown in Fig.~\ref{fig:fig3}(e). Although our data suffers from finite-size effects due to the curvature of the dispersion \cite{suppl}, we see a tendency for the velocity differences to decrease on approaching the transition from dSC to dSC$_\text{g}$+AFM. This is in line with the field theory prediction that at long distances, all three velocities become equal at this transition \cite{balents1998nodal}.

{\it Low energy theory}\,---\,
The transition from the nodal dSC to the dSC$_\text{g}$+AFM is of particular interest since it hosts gapless nodal fermions. 
The corresponding low energy theory was discussed in Ref.~\cite{balents1998nodal}.
In the following we will show that it can be further mapped to the  HGN theory. 
The nodal dSC phase has nodes located at $(\pm\frac{\pi}{2}, \pm\frac{\pi}{2})$, as shown in Fig.~\ref{fig:fig3}. 
These four nodes can be divided into two pairs, forming two four-component Dirac fermions.
At this transition, in addition to the nodal fermions, we also have gapless AFM modes $\vec{N}$, which couple linearly to the corresponding Dirac fermion bilinear. After a few basis transformations, while ignoring the difference between the three  velocities $\upsilon_F, \upsilon_\Delta, \upsilon_s$, and rescaling them to unity, one arrives at the following Lagrangian~\cite{balents1998nodal}:
\begin{equation}
\mathcal{L} = \bar{\Psi} \slashed{\partial} \Psi + \frac{1}{2}(\partial_\mu \vec{N})^2 +  u (\vec{N}^2)^2 + g\vec{N}\cdot(\Psi^\dagger \tau^y \vec{\sigma}\sigma^y\Psi^\dagger + \text{h.c.}), 
\end{equation}
where $\slashed{\partial}=\gamma^\mu\partial_\mu$, and the Pauli matrices $\vec{\tau}$ act in the particle-hole space of the original microscopic fermions $c$.
The $4-\epsilon$ RG calculations predict that the difference between velocities eventually flows to zero, so that the above isotropic description is valid for a continuous transition.
The isotropic free Dirac fermions have an $O(8)$ symmetry, which can be made manifest by employing Majorana basis ($\Psi=\eta_1+\i\eta_2$). We will now exploit this $O(8)$ symmetry to transform the critical theory into a well-known form. The coupling term in the Majorana basis becomes $g\vec{N}\cdot(\eta^T\tau^y\vec{\Sigma}\eta)$ with $\vec{\Sigma}=(\sigma^z\rho^x,\rho^z,-\sigma^x\rho^x)$, where the Pauli matrices $\rho^{x,y,z}$ act in the Majorana space $(\eta_1, \eta_2)$. With an orthogonal transformation $O=\frac{\i}{\sqrt{2}}(\sigma^y\rho^z-\rho^y)\in O(8)$ of the free theory, the coupling can be transformed into $g\vec{N}\cdot(\eta'^T\tau^y\vec{\Sigma}'\eta')$ with $\eta' = O \eta$ and $\vec{\Sigma}'=(\sigma^x,-\sigma^y\rho^y,\sigma^z)$, which if written in terms of complex fermions, is nothing but the standard HGN coupling $g\vec{N}\cdot(\Psi'^\dagger \tau^y\vec{\sigma}\Psi')$ where $\Psi' = \eta'_1 + i \eta'_2$. Therefore, the low energy theory is equivalent to the HGN model, and the transition should also belong to the chiral Heisenberg universality class with two four-component Dirac fermions~\cite{Rosenstein1993, Herbut06,Herbut09,Assaad13,Janssen2014,
Toldin14,Otsuka16,Zerf2017,Gracey2018,
Buividovich2018,Lang2019}. The above field theory arguments also apply to a recent numerical QMC study of mean-field Hamiltonian of dSC to AFM transition~\cite{otsuka2020dirac}. The charge $U(1)$ symmetry is \textit{explicitly} broken in their model, and therefore it doesn't support a phase where the superconductor results from spontaneous symmetry breaking, or an AFM phase which preserves charge $U(1)$. However,  HGN is still the correct effective description of the spin-$SU(2)$ breaking \textit{transition} seen in their model. This is because the charge-$U(1)$ doesn't play an important role across this transition from RG perspective, and our above analysis applies. The absence of charge $U(1)$ also implies that in contrast to our model, there will not be a finite-T Berezinskii-Kosterlitz-Thouless (BKT) transition. Although the complexity of our model limits us to 20 $\times$ 20 system size, the critical exponents we found, namely $\nu \approx 0.99$ and $\eta_m \approx 0.55$, are consistent with works which study the same universality class upto 40 $\times$ 40 size \cite{Otsuka16, otsuka2020dirac}. The limited size also limits our ability to perform a more detailed analysis of finite size corrections. It's worth noting that the value of $\eta_m$ reported in previous numerical works differ from each other considerably, and lies in a window ranging from $0.45$ to $1.2$, which likely signals strong finite-size corrections for this exponent even for larger system sizes.
%

{\it Discussion and conclusion}\,---\,
Very broadly, the structure of our model is in the similar spirit as Refs.\cite{Berg12, Schattner2016, Berg2019,
Xu2019,Li2019}, where a desired ordered phase is obtained by coupling the corresponding fermionic bilinear to a fluctuating bosonic field, and then tuning the kinetic energy of the boson to obtain an order-disorder quantum critical point. The novelty of our model is that it allows one to access a spontaneously symmetry broken nodal SC phase, and its competition with AFM, a scenario realized in a large class of materials \cite{Norman2011, Dagotto94, Lee06_rev, steglich1984heavy,Stewart1984, Amato1997,Pfleiderer2009,Pagliuso2001,maple_rev2010,Steglich79,Joynt2002,Knebel2004,
Park2006,Yuan2003,Mito2003,Aeppli1987, stewart2011superconductivity, si2016high, Dai2015, Pratt2009, Nandi2010, vollhardt2013superfluid, balents2020superconductivity}, and especially pressure tuned SC to AFM transition in layered organic Mott insulators \cite{Kagawa05,Kurosaki2005,Lefebvre00,Arai2001,
Belin1998,DeSoto1995,Mayaffre1995,
Kanoda1996,Jeerome1991}. Although our Hamiltonian involves terms that are not conventional, owing to the usual notions of universality of low-energy effective theory, we expect aspects such as the nature of quantum criticality as well as the topology of the phase diagram to correctly capture more realistic Hamiltonians (which typically suffer from the sign problem), as long as the symmetries and the filling match with our model. At the same time, we do not attempt to provide insight into questions such as the shape of the phase boundaries, effects related to geometrical frustration, or even the underlying mechanism for superconductivity in the aforementioned materials.

The competition between d-wave superconductivity and AFM is also explored in Monte Carlo studies of various multiband models~\cite{Berg2016, dumitrescu2016, li2017nature, lederer2017superconductivity,
Wang2017,christensen2020modeling}. In contrast to our model, in Refs.\cite{Berg2016, dumitrescu2016, li2017nature, lederer2017superconductivity, Wang2017,christensen2020modeling}, 
the pairing order-parameter is in fact onsite, and the superconductivity is not nodal. 
We also note an interesting proposal for a model of competing AFM and nodal dSC phases, Ref.\cite{assaad1996quantum}. In this purely electronic model, the value of the d-wave order-parameter obtained on the largest system size was rather small.
and it might be interesting to revisit it.

Our model allows for adding orbital
and Zeeman magnetic fields of arbitrary strength \textit{without} introducing  sign-problem. It will be very interesting to study the destruction of dSC as either of these fields are ramped up. Another interesting question is to explore the possibility of destroying the dSC phase by proliferating \textit{double} vortices, while keeping a single vortex gapped. This will lead to a fractionalized phase with topological order where nodal spinons are coupled to a $Z_2$ gauge field. 


In summary, we constructed a sign-problem-free repulsive Hubbard model which hosts competing AFM and nodal d-wave phases. We found three different phases by tuning a single parameter $K/J$ that controls the fluctuations of the bosonic fields: a nodal dSC phase at small $K/J$, an AFM phase at large $K/J$, and an intermediate phase with coexisting gapped d-wave and AFM orders. The phase transition between the coexistence phase and the AFM phase is a conventional 3D XY transition, while  that between the coexistence phase and the nodal d-wave also appears to be continuous, and belongs to the 2+1-D HGN universality class. 

\textit{Note added:} During the completion of this manuscript, we became aware of a recent preprint arXiv:2009.04685~\cite{otsuka2020dirac}, which studies a \textit{mean-field} Hamiltonian of nodal dSC with repulsive Hubbard-$U$. The relation to this work is discussed above in the Sec. \textit{Low energy theory}.

\begin{acknowledgments}
We thank Fakher Assaad, John McGreevy for useful discussions, and Dan Arovas, Rafael Fernandes for comments on the draft. TG acknowledges support by the National Science Foundation under Grant No. DMR-1752417, and by an Alfred P. Sloan Research Fellowship. This work used the Extreme Science and Engineering Discovery Environment (XSEDE)~\cite{xsede}, which is supported by National Science Foundation grant number ACI-1548562.
\end{acknowledgments}

\bibliographystyle{apsrev4-1}
\bibliography{corr}

\begin{thebibliography}{75}%
\makeatletter
\providecommand \@ifxundefined [1]{%
 \@ifx{#1\undefined}
}%
\providecommand \@ifnum [1]{%
 \ifnum #1\expandafter \@firstoftwo
 \else \expandafter \@secondoftwo
 \fi
}%
\providecommand \@ifx [1]{%
 \ifx #1\expandafter \@firstoftwo
 \else \expandafter \@secondoftwo
 \fi
}%
\providecommand \natexlab [1]{#1}%
\providecommand \enquote  [1]{``#1''}%
\providecommand \bibnamefont  [1]{#1}%
\providecommand \bibfnamefont [1]{#1}%
\providecommand \citenamefont [1]{#1}%
\providecommand \href@noop [0]{\@secondoftwo}%
\providecommand \href [0]{\begingroup \@sanitize@url \@href}%
\providecommand \@href[1]{\@@startlink{#1}\@@href}%
\providecommand \@@href[1]{\endgroup#1\@@endlink}%
\providecommand \@sanitize@url [0]{\catcode `\\12\catcode `\$12\catcode
  `\&12\catcode `\#12\catcode `\^12\catcode `\_12\catcode `\%12\relax}%
\providecommand \@@startlink[1]{}%
\providecommand \@@endlink[0]{}%
\providecommand \url  [0]{\begingroup\@sanitize@url \@url }%
\providecommand \@url [1]{\endgroup\@href {#1}{\urlprefix }}%
\providecommand \urlprefix  [0]{URL }%
\providecommand \Eprint [0]{\href }%
\providecommand \doibase [0]{http://dx.doi.org/}%
\providecommand \selectlanguage [0]{\@gobble}%
\providecommand \bibinfo  [0]{\@secondoftwo}%
\providecommand \bibfield  [0]{\@secondoftwo}%
\providecommand \translation [1]{[#1]}%
\providecommand \BibitemOpen [0]{}%
\providecommand \bibitemStop [0]{}%
\providecommand \bibitemNoStop [0]{.\EOS\space}%
\providecommand \EOS [0]{\spacefactor3000\relax}%
\providecommand \BibitemShut  [1]{\csname bibitem#1\endcsname}%
\let\auto@bib@innerbib\@empty
\bibitem [{\citenamefont {Norman}(2011)}]{Norman2011}%
  \BibitemOpen
  \bibfield  {author} {\bibinfo {author} {\bibfnamefont {M.~R.}\ \bibnamefont
  {Norman}},\ }\href {\doibase 10.1126/science.1200181} {\bibfield  {journal}
  {\bibinfo  {journal} {Science}\ }\textbf {\bibinfo {volume} {332}},\ \bibinfo
  {pages} {196} (\bibinfo {year} {2011})}\BibitemShut {NoStop}%
\bibitem [{\citenamefont {Dagotto}(1994)}]{Dagotto94}%
  \BibitemOpen
  \bibfield  {author} {\bibinfo {author} {\bibfnamefont {E.}~\bibnamefont
  {Dagotto}},\ }\href {\doibase 10.1103/RevModPhys.66.763} {\bibfield
  {journal} {\bibinfo  {journal} {Rev. Mod. Phys.}\ }\textbf {\bibinfo {volume}
  {66}},\ \bibinfo {pages} {763} (\bibinfo {year} {1994})}\BibitemShut
  {NoStop}%
\bibitem [{\citenamefont {Lee}\ \emph {et~al.}(2006)\citenamefont {Lee},
  \citenamefont {Nagaosa},\ and\ \citenamefont {Wen}}]{Lee06_rev}%
  \BibitemOpen
  \bibfield  {author} {\bibinfo {author} {\bibfnamefont {P.~A.}\ \bibnamefont
  {Lee}}, \bibinfo {author} {\bibfnamefont {N.}~\bibnamefont {Nagaosa}}, \ and\
  \bibinfo {author} {\bibfnamefont {X.-G.}\ \bibnamefont {Wen}},\ }\href
  {\doibase 10.1103/RevModPhys.78.17} {\bibfield  {journal} {\bibinfo
  {journal} {Rev. Mod. Phys.}\ }\textbf {\bibinfo {volume} {78}},\ \bibinfo
  {pages} {17} (\bibinfo {year} {2006})}\BibitemShut {NoStop}%
\bibitem [{\citenamefont {Steglich}\ \emph {et~al.}(1984)\citenamefont
  {Steglich}, \citenamefont {Bredl}, \citenamefont {Lieke}, \citenamefont
  {Rauchschwalbe},\ and\ \citenamefont {Sparn}}]{steglich1984heavy}%
  \BibitemOpen
  \bibfield  {author} {\bibinfo {author} {\bibfnamefont {F.}~\bibnamefont
  {Steglich}}, \bibinfo {author} {\bibfnamefont {C.}~\bibnamefont {Bredl}},
  \bibinfo {author} {\bibfnamefont {W.}~\bibnamefont {Lieke}}, \bibinfo
  {author} {\bibfnamefont {U.}~\bibnamefont {Rauchschwalbe}}, \ and\ \bibinfo
  {author} {\bibfnamefont {G.}~\bibnamefont {Sparn}},\ }\href {\doibase
  https://doi.org/10.1016/0378-4363(84)90148-7} {\bibfield  {journal} {\bibinfo
   {journal} {Physica B+C}\ }\textbf {\bibinfo {volume} {126}},\ \bibinfo
  {pages} {82 } (\bibinfo {year} {1984})}\BibitemShut {NoStop}%
\bibitem [{\citenamefont {Stewart}(1984)}]{Stewart1984}%
  \BibitemOpen
  \bibfield  {author} {\bibinfo {author} {\bibfnamefont {G.~R.}\ \bibnamefont
  {Stewart}},\ }\href {\doibase 10.1103/RevModPhys.56.755} {\bibfield
  {journal} {\bibinfo  {journal} {Rev. Mod. Phys.}\ }\textbf {\bibinfo {volume}
  {56}},\ \bibinfo {pages} {755} (\bibinfo {year} {1984})}\BibitemShut
  {NoStop}%
\bibitem [{\citenamefont {Amato}(1997)}]{Amato1997}%
  \BibitemOpen
  \bibfield  {author} {\bibinfo {author} {\bibfnamefont {A.}~\bibnamefont
  {Amato}},\ }\href {\doibase 10.1103/RevModPhys.69.1119} {\bibfield  {journal}
  {\bibinfo  {journal} {Rev. Mod. Phys.}\ }\textbf {\bibinfo {volume} {69}},\
  \bibinfo {pages} {1119} (\bibinfo {year} {1997})}\BibitemShut {NoStop}%
\bibitem [{\citenamefont {Pfleiderer}(2009)}]{Pfleiderer2009}%
  \BibitemOpen
  \bibfield  {author} {\bibinfo {author} {\bibfnamefont {C.}~\bibnamefont
  {Pfleiderer}},\ }\href {\doibase 10.1103/RevModPhys.81.1551} {\bibfield
  {journal} {\bibinfo  {journal} {Rev. Mod. Phys.}\ }\textbf {\bibinfo {volume}
  {81}},\ \bibinfo {pages} {1551} (\bibinfo {year} {2009})}\BibitemShut
  {NoStop}%
\bibitem [{\citenamefont {Pagliuso}\ \emph {et~al.}(2001)\citenamefont
  {Pagliuso}, \citenamefont {Petrovic}, \citenamefont {Movshovich},
  \citenamefont {Hall}, \citenamefont {Hundley}, \citenamefont {Sarrao},
  \citenamefont {Thompson},\ and\ \citenamefont {Fisk}}]{Pagliuso2001}%
  \BibitemOpen
  \bibfield  {author} {\bibinfo {author} {\bibfnamefont {P.~G.}\ \bibnamefont
  {Pagliuso}}, \bibinfo {author} {\bibfnamefont {C.}~\bibnamefont {Petrovic}},
  \bibinfo {author} {\bibfnamefont {R.}~\bibnamefont {Movshovich}}, \bibinfo
  {author} {\bibfnamefont {D.}~\bibnamefont {Hall}}, \bibinfo {author}
  {\bibfnamefont {M.~F.}\ \bibnamefont {Hundley}}, \bibinfo {author}
  {\bibfnamefont {J.~L.}\ \bibnamefont {Sarrao}}, \bibinfo {author}
  {\bibfnamefont {J.~D.}\ \bibnamefont {Thompson}}, \ and\ \bibinfo {author}
  {\bibfnamefont {Z.}~\bibnamefont {Fisk}},\ }\href {\doibase
  10.1103/PhysRevB.64.100503} {\bibfield  {journal} {\bibinfo  {journal} {Phys.
  Rev. B}\ }\textbf {\bibinfo {volume} {64}},\ \bibinfo {pages} {100503}
  (\bibinfo {year} {2001})}\BibitemShut {NoStop}%
\bibitem [{\citenamefont {Brian~Maple}\ \emph {et~al.}(2010)\citenamefont
  {Brian~Maple}, \citenamefont {Baumbach}, \citenamefont {Butch}, \citenamefont
  {Hamlin},\ and\ \citenamefont {Janoschek}}]{maple_rev2010}%
  \BibitemOpen
  \bibfield  {author} {\bibinfo {author} {\bibfnamefont {M.}~\bibnamefont
  {Brian~Maple}}, \bibinfo {author} {\bibfnamefont {R.~E.}\ \bibnamefont
  {Baumbach}}, \bibinfo {author} {\bibfnamefont {N.~P.}\ \bibnamefont {Butch}},
  \bibinfo {author} {\bibfnamefont {J.~J.}\ \bibnamefont {Hamlin}}, \ and\
  \bibinfo {author} {\bibfnamefont {M.}~\bibnamefont {Janoschek}},\ }\href
  {\doibase 10.1007/s10909-010-0212-5} {\bibfield  {journal} {\bibinfo
  {journal} {Journal of Low Temperature Physics}\ }\textbf {\bibinfo {volume}
  {161}},\ \bibinfo {pages} {4} (\bibinfo {year} {2010})}\BibitemShut {NoStop}%
\bibitem [{\citenamefont {Steglich}\ \emph {et~al.}(1979)\citenamefont
  {Steglich}, \citenamefont {Aarts}, \citenamefont {Bredl}, \citenamefont
  {Lieke}, \citenamefont {Meschede}, \citenamefont {Franz},\ and\ \citenamefont
  {Sch\"afer}}]{Steglich79}%
  \BibitemOpen
  \bibfield  {author} {\bibinfo {author} {\bibfnamefont {F.}~\bibnamefont
  {Steglich}}, \bibinfo {author} {\bibfnamefont {J.}~\bibnamefont {Aarts}},
  \bibinfo {author} {\bibfnamefont {C.~D.}\ \bibnamefont {Bredl}}, \bibinfo
  {author} {\bibfnamefont {W.}~\bibnamefont {Lieke}}, \bibinfo {author}
  {\bibfnamefont {D.}~\bibnamefont {Meschede}}, \bibinfo {author}
  {\bibfnamefont {W.}~\bibnamefont {Franz}}, \ and\ \bibinfo {author}
  {\bibfnamefont {H.}~\bibnamefont {Sch\"afer}},\ }\href {\doibase
  10.1103/PhysRevLett.43.1892} {\bibfield  {journal} {\bibinfo  {journal}
  {Phys. Rev. Lett.}\ }\textbf {\bibinfo {volume} {43}},\ \bibinfo {pages}
  {1892} (\bibinfo {year} {1979})}\BibitemShut {NoStop}%
\bibitem [{\citenamefont {Joynt}\ and\ \citenamefont
  {Taillefer}(2002)}]{Joynt2002}%
  \BibitemOpen
  \bibfield  {author} {\bibinfo {author} {\bibfnamefont {R.}~\bibnamefont
  {Joynt}}\ and\ \bibinfo {author} {\bibfnamefont {L.}~\bibnamefont
  {Taillefer}},\ }\href {\doibase 10.1103/RevModPhys.74.235} {\bibfield
  {journal} {\bibinfo  {journal} {Rev. Mod. Phys.}\ }\textbf {\bibinfo {volume}
  {74}},\ \bibinfo {pages} {235} (\bibinfo {year} {2002})}\BibitemShut
  {NoStop}%
\bibitem [{\citenamefont {Knebel}\ \emph {et~al.}(2004)\citenamefont {Knebel},
  \citenamefont {M{\'{e}}asson}, \citenamefont {Salce}, \citenamefont {Aoki},
  \citenamefont {Braithwaite}, \citenamefont {Brison},\ and\ \citenamefont
  {Flouquet}}]{Knebel2004}%
  \BibitemOpen
  \bibfield  {author} {\bibinfo {author} {\bibfnamefont {G.}~\bibnamefont
  {Knebel}}, \bibinfo {author} {\bibfnamefont {M.-A.}\ \bibnamefont
  {M{\'{e}}asson}}, \bibinfo {author} {\bibfnamefont {B.}~\bibnamefont
  {Salce}}, \bibinfo {author} {\bibfnamefont {D.}~\bibnamefont {Aoki}},
  \bibinfo {author} {\bibfnamefont {D.}~\bibnamefont {Braithwaite}}, \bibinfo
  {author} {\bibfnamefont {J.~P.}\ \bibnamefont {Brison}}, \ and\ \bibinfo
  {author} {\bibfnamefont {J.}~\bibnamefont {Flouquet}},\ }\href {\doibase
  10.1088/0953-8984/16/49/008} {\bibfield  {journal} {\bibinfo  {journal}
  {Journal of Physics: Condensed Matter}\ }\textbf {\bibinfo {volume} {16}},\
  \bibinfo {pages} {8905} (\bibinfo {year} {2004})}\BibitemShut {NoStop}%
\bibitem [{\citenamefont {Park}\ \emph {et~al.}(2006)\citenamefont {Park},
  \citenamefont {Ronning}, \citenamefont {Yuan}, \citenamefont {Salamon},
  \citenamefont {Movshovich}, \citenamefont {Sarrao},\ and\ \citenamefont
  {Thompson}}]{Park2006}%
  \BibitemOpen
  \bibfield  {author} {\bibinfo {author} {\bibfnamefont {T.}~\bibnamefont
  {Park}}, \bibinfo {author} {\bibfnamefont {F.}~\bibnamefont {Ronning}},
  \bibinfo {author} {\bibfnamefont {H.~Q.}\ \bibnamefont {Yuan}}, \bibinfo
  {author} {\bibfnamefont {M.~B.}\ \bibnamefont {Salamon}}, \bibinfo {author}
  {\bibfnamefont {R.}~\bibnamefont {Movshovich}}, \bibinfo {author}
  {\bibfnamefont {J.~L.}\ \bibnamefont {Sarrao}}, \ and\ \bibinfo {author}
  {\bibfnamefont {J.~D.}\ \bibnamefont {Thompson}},\ }\href {\doibase
  10.1038/nature04571} {\bibfield  {journal} {\bibinfo  {journal} {Nature}\
  }\textbf {\bibinfo {volume} {440}},\ \bibinfo {pages} {65} (\bibinfo {year}
  {2006})}\BibitemShut {NoStop}%
\bibitem [{\citenamefont {Yuan}\ \emph {et~al.}(2003)\citenamefont {Yuan},
  \citenamefont {Grosche}, \citenamefont {Deppe}, \citenamefont {Geibel},
  \citenamefont {Sparn},\ and\ \citenamefont {Steglich}}]{Yuan2003}%
  \BibitemOpen
  \bibfield  {author} {\bibinfo {author} {\bibfnamefont {H.~Q.}\ \bibnamefont
  {Yuan}}, \bibinfo {author} {\bibfnamefont {F.~M.}\ \bibnamefont {Grosche}},
  \bibinfo {author} {\bibfnamefont {M.}~\bibnamefont {Deppe}}, \bibinfo
  {author} {\bibfnamefont {C.}~\bibnamefont {Geibel}}, \bibinfo {author}
  {\bibfnamefont {G.}~\bibnamefont {Sparn}}, \ and\ \bibinfo {author}
  {\bibfnamefont {F.}~\bibnamefont {Steglich}},\ }\href {\doibase
  10.1126/science.1091648} {\bibfield  {journal} {\bibinfo  {journal}
  {Science}\ }\textbf {\bibinfo {volume} {302}},\ \bibinfo {pages} {2104}
  (\bibinfo {year} {2003})}\BibitemShut {NoStop}%
\bibitem [{\citenamefont {Mito}\ \emph {et~al.}(2003)\citenamefont {Mito},
  \citenamefont {Kawasaki}, \citenamefont {Kawasaki}, \citenamefont {Zheng},
  \citenamefont {Kitaoka}, \citenamefont {Aoki}, \citenamefont {Haga},\ and\
  \citenamefont {\ifmmode~\bar{O}\else \={O}\fi{}nuki}}]{Mito2003}%
  \BibitemOpen
  \bibfield  {author} {\bibinfo {author} {\bibfnamefont {T.}~\bibnamefont
  {Mito}}, \bibinfo {author} {\bibfnamefont {S.}~\bibnamefont {Kawasaki}},
  \bibinfo {author} {\bibfnamefont {Y.}~\bibnamefont {Kawasaki}}, \bibinfo
  {author} {\bibfnamefont {G.~q.}\ \bibnamefont {Zheng}}, \bibinfo {author}
  {\bibfnamefont {Y.}~\bibnamefont {Kitaoka}}, \bibinfo {author} {\bibfnamefont
  {D.}~\bibnamefont {Aoki}}, \bibinfo {author} {\bibfnamefont {Y.~.}\
  \bibnamefont {Haga}}, \ and\ \bibinfo {author} {\bibfnamefont
  {Y.}~\bibnamefont {\ifmmode~\bar{O}\else \={O}\fi{}nuki}},\ }\href {\doibase
  10.1103/PhysRevLett.90.077004} {\bibfield  {journal} {\bibinfo  {journal}
  {Phys. Rev. Lett.}\ }\textbf {\bibinfo {volume} {90}},\ \bibinfo {pages}
  {077004} (\bibinfo {year} {2003})}\BibitemShut {NoStop}%
\bibitem [{\citenamefont {Aeppli}\ \emph {et~al.}(1987)\citenamefont {Aeppli},
  \citenamefont {Goldman}, \citenamefont {Shirane}, \citenamefont {Bucher},\
  and\ \citenamefont {Lux-Steiner}}]{Aeppli1987}%
  \BibitemOpen
  \bibfield  {author} {\bibinfo {author} {\bibfnamefont {G.}~\bibnamefont
  {Aeppli}}, \bibinfo {author} {\bibfnamefont {A.}~\bibnamefont {Goldman}},
  \bibinfo {author} {\bibfnamefont {G.}~\bibnamefont {Shirane}}, \bibinfo
  {author} {\bibfnamefont {E.}~\bibnamefont {Bucher}}, \ and\ \bibinfo {author}
  {\bibfnamefont {M.-C.}\ \bibnamefont {Lux-Steiner}},\ }\href {\doibase
  10.1103/PhysRevLett.58.808} {\bibfield  {journal} {\bibinfo  {journal} {Phys.
  Rev. Lett.}\ }\textbf {\bibinfo {volume} {58}},\ \bibinfo {pages} {808}
  (\bibinfo {year} {1987})}\BibitemShut {NoStop}%
\bibitem [{\citenamefont {Kagawa}\ \emph {et~al.}(2005)\citenamefont {Kagawa},
  \citenamefont {Miyagawa},\ and\ \citenamefont {Kanoda}}]{Kagawa05}%
  \BibitemOpen
  \bibfield  {author} {\bibinfo {author} {\bibfnamefont {F.}~\bibnamefont
  {Kagawa}}, \bibinfo {author} {\bibfnamefont {K.}~\bibnamefont {Miyagawa}}, \
  and\ \bibinfo {author} {\bibfnamefont {K.}~\bibnamefont {Kanoda}},\ }\href
  {http://dx.doi.org/10.1038/nature03806} {\bibfield  {journal} {\bibinfo
  {journal} {Nature}\ }\textbf {\bibinfo {volume} {436}},\ \bibinfo {pages}
  {534} (\bibinfo {year} {2005})}\BibitemShut {NoStop}%
\bibitem [{\citenamefont {Kurosaki}\ \emph {et~al.}(2005)\citenamefont
  {Kurosaki}, \citenamefont {Shimizu}, \citenamefont {Miyagawa}, \citenamefont
  {Kanoda},\ and\ \citenamefont {Saito}}]{Kurosaki2005}%
  \BibitemOpen
  \bibfield  {author} {\bibinfo {author} {\bibfnamefont {Y.}~\bibnamefont
  {Kurosaki}}, \bibinfo {author} {\bibfnamefont {Y.}~\bibnamefont {Shimizu}},
  \bibinfo {author} {\bibfnamefont {K.}~\bibnamefont {Miyagawa}}, \bibinfo
  {author} {\bibfnamefont {K.}~\bibnamefont {Kanoda}}, \ and\ \bibinfo {author}
  {\bibfnamefont {G.}~\bibnamefont {Saito}},\ }\href {\doibase
  10.1103/PhysRevLett.95.177001} {\bibfield  {journal} {\bibinfo  {journal}
  {Phys. Rev. Lett.}\ }\textbf {\bibinfo {volume} {95}},\ \bibinfo {pages}
  {177001} (\bibinfo {year} {2005})}\BibitemShut {NoStop}%
\bibitem [{\citenamefont {Lefebvre}\ \emph {et~al.}(2000)\citenamefont
  {Lefebvre}, \citenamefont {Wzietek}, \citenamefont {Brown}, \citenamefont
  {Bourbonnais}, \citenamefont {J\'erome}, \citenamefont {M\'ezi\`ere},
  \citenamefont {Fourmigu\'e},\ and\ \citenamefont {Batail}}]{Lefebvre00}%
  \BibitemOpen
  \bibfield  {author} {\bibinfo {author} {\bibfnamefont {S.}~\bibnamefont
  {Lefebvre}}, \bibinfo {author} {\bibfnamefont {P.}~\bibnamefont {Wzietek}},
  \bibinfo {author} {\bibfnamefont {S.}~\bibnamefont {Brown}}, \bibinfo
  {author} {\bibfnamefont {C.}~\bibnamefont {Bourbonnais}}, \bibinfo {author}
  {\bibfnamefont {D.}~\bibnamefont {J\'erome}}, \bibinfo {author}
  {\bibfnamefont {C.}~\bibnamefont {M\'ezi\`ere}}, \bibinfo {author}
  {\bibfnamefont {M.}~\bibnamefont {Fourmigu\'e}}, \ and\ \bibinfo {author}
  {\bibfnamefont {P.}~\bibnamefont {Batail}},\ }\href {\doibase
  10.1103/PhysRevLett.85.5420} {\bibfield  {journal} {\bibinfo  {journal}
  {Phys. Rev. Lett.}\ }\textbf {\bibinfo {volume} {85}},\ \bibinfo {pages}
  {5420} (\bibinfo {year} {2000})}\BibitemShut {NoStop}%
\bibitem [{\citenamefont {Arai}\ \emph {et~al.}(2001)\citenamefont {Arai},
  \citenamefont {Ichimura}, \citenamefont {Nomura}, \citenamefont {Takasaki},
  \citenamefont {Yamada}, \citenamefont {Nakatsuji},\ and\ \citenamefont
  {Anzai}}]{Arai2001}%
  \BibitemOpen
  \bibfield  {author} {\bibinfo {author} {\bibfnamefont {T.}~\bibnamefont
  {Arai}}, \bibinfo {author} {\bibfnamefont {K.}~\bibnamefont {Ichimura}},
  \bibinfo {author} {\bibfnamefont {K.}~\bibnamefont {Nomura}}, \bibinfo
  {author} {\bibfnamefont {S.}~\bibnamefont {Takasaki}}, \bibinfo {author}
  {\bibfnamefont {J.}~\bibnamefont {Yamada}}, \bibinfo {author} {\bibfnamefont
  {S.}~\bibnamefont {Nakatsuji}}, \ and\ \bibinfo {author} {\bibfnamefont
  {H.}~\bibnamefont {Anzai}},\ }\href {\doibase 10.1103/PhysRevB.63.104518}
  {\bibfield  {journal} {\bibinfo  {journal} {Phys. Rev. B}\ }\textbf {\bibinfo
  {volume} {63}},\ \bibinfo {pages} {104518} (\bibinfo {year}
  {2001})}\BibitemShut {NoStop}%
\bibitem [{\citenamefont {Belin}\ \emph {et~al.}(1998)\citenamefont {Belin},
  \citenamefont {Behnia},\ and\ \citenamefont {Deluzet}}]{Belin1998}%
  \BibitemOpen
  \bibfield  {author} {\bibinfo {author} {\bibfnamefont {S.}~\bibnamefont
  {Belin}}, \bibinfo {author} {\bibfnamefont {K.}~\bibnamefont {Behnia}}, \
  and\ \bibinfo {author} {\bibfnamefont {A.}~\bibnamefont {Deluzet}},\ }\href
  {\doibase 10.1103/PhysRevLett.81.4728} {\bibfield  {journal} {\bibinfo
  {journal} {Phys. Rev. Lett.}\ }\textbf {\bibinfo {volume} {81}},\ \bibinfo
  {pages} {4728} (\bibinfo {year} {1998})}\BibitemShut {NoStop}%
\bibitem [{\citenamefont {De~Soto}\ \emph {et~al.}(1995)\citenamefont
  {De~Soto}, \citenamefont {Slichter}, \citenamefont {Kini}, \citenamefont
  {Wang}, \citenamefont {Geiser},\ and\ \citenamefont {Williams}}]{DeSoto1995}%
  \BibitemOpen
  \bibfield  {author} {\bibinfo {author} {\bibfnamefont {S.~M.}\ \bibnamefont
  {De~Soto}}, \bibinfo {author} {\bibfnamefont {C.~P.}\ \bibnamefont
  {Slichter}}, \bibinfo {author} {\bibfnamefont {A.~M.}\ \bibnamefont {Kini}},
  \bibinfo {author} {\bibfnamefont {H.~H.}\ \bibnamefont {Wang}}, \bibinfo
  {author} {\bibfnamefont {U.}~\bibnamefont {Geiser}}, \ and\ \bibinfo {author}
  {\bibfnamefont {J.~M.}\ \bibnamefont {Williams}},\ }\href {\doibase
  10.1103/PhysRevB.52.10364} {\bibfield  {journal} {\bibinfo  {journal} {Phys.
  Rev. B}\ }\textbf {\bibinfo {volume} {52}},\ \bibinfo {pages} {10364}
  (\bibinfo {year} {1995})}\BibitemShut {NoStop}%
\bibitem [{\citenamefont {Mayaffre}\ \emph {et~al.}(1995)\citenamefont
  {Mayaffre}, \citenamefont {Wzietek}, \citenamefont {J\'erome}, \citenamefont
  {Lenoir},\ and\ \citenamefont {Batail}}]{Mayaffre1995}%
  \BibitemOpen
  \bibfield  {author} {\bibinfo {author} {\bibfnamefont {H.}~\bibnamefont
  {Mayaffre}}, \bibinfo {author} {\bibfnamefont {P.}~\bibnamefont {Wzietek}},
  \bibinfo {author} {\bibfnamefont {D.}~\bibnamefont {J\'erome}}, \bibinfo
  {author} {\bibfnamefont {C.}~\bibnamefont {Lenoir}}, \ and\ \bibinfo {author}
  {\bibfnamefont {P.}~\bibnamefont {Batail}},\ }\href {\doibase
  10.1103/PhysRevLett.75.4122} {\bibfield  {journal} {\bibinfo  {journal}
  {Phys. Rev. Lett.}\ }\textbf {\bibinfo {volume} {75}},\ \bibinfo {pages}
  {4122} (\bibinfo {year} {1995})}\BibitemShut {NoStop}%
\bibitem [{\citenamefont {Kanoda}\ \emph {et~al.}(1996)\citenamefont {Kanoda},
  \citenamefont {Miyagawa}, \citenamefont {Kawamoto},\ and\ \citenamefont
  {Nakazawa}}]{Kanoda1996}%
  \BibitemOpen
  \bibfield  {author} {\bibinfo {author} {\bibfnamefont {K.}~\bibnamefont
  {Kanoda}}, \bibinfo {author} {\bibfnamefont {K.}~\bibnamefont {Miyagawa}},
  \bibinfo {author} {\bibfnamefont {A.}~\bibnamefont {Kawamoto}}, \ and\
  \bibinfo {author} {\bibfnamefont {Y.}~\bibnamefont {Nakazawa}},\ }\href
  {\doibase 10.1103/PhysRevB.54.76} {\bibfield  {journal} {\bibinfo  {journal}
  {Phys. Rev. B}\ }\textbf {\bibinfo {volume} {54}},\ \bibinfo {pages} {76}
  (\bibinfo {year} {1996})}\BibitemShut {NoStop}%
\bibitem [{\citenamefont {J{\'E}ROME}(1991)}]{Jeerome1991}%
  \BibitemOpen
  \bibfield  {author} {\bibinfo {author} {\bibfnamefont {D.}~\bibnamefont
  {J{\'E}ROME}},\ }\href {\doibase 10.1126/science.252.5012.1509} {\bibfield
  {journal} {\bibinfo  {journal} {Science}\ }\textbf {\bibinfo {volume}
  {252}},\ \bibinfo {pages} {1509} (\bibinfo {year} {1991})}\BibitemShut
  {NoStop}%
\bibitem [{\citenamefont {Stewart}(2011)}]{stewart2011superconductivity}%
  \BibitemOpen
  \bibfield  {author} {\bibinfo {author} {\bibfnamefont {G.}~\bibnamefont
  {Stewart}},\ }\href@noop {} {\bibfield  {journal} {\bibinfo  {journal}
  {Reviews of Modern Physics}\ }\textbf {\bibinfo {volume} {83}},\ \bibinfo
  {pages} {1589} (\bibinfo {year} {2011})}\BibitemShut {NoStop}%
\bibitem [{\citenamefont {Si}\ \emph {et~al.}(2016)\citenamefont {Si},
  \citenamefont {Yu},\ and\ \citenamefont {Abrahams}}]{si2016high}%
  \BibitemOpen
  \bibfield  {author} {\bibinfo {author} {\bibfnamefont {Q.}~\bibnamefont
  {Si}}, \bibinfo {author} {\bibfnamefont {R.}~\bibnamefont {Yu}}, \ and\
  \bibinfo {author} {\bibfnamefont {E.}~\bibnamefont {Abrahams}},\ }\href
  {\doibase 10.1038/natrevmats.2016.17} {\bibfield  {journal} {\bibinfo
  {journal} {Nature Reviews Materials}\ }\textbf {\bibinfo {volume} {1}},\
  \bibinfo {pages} {16017} (\bibinfo {year} {2016})}\BibitemShut {NoStop}%
\bibitem [{\citenamefont {Dai}(2015)}]{Dai2015}%
  \BibitemOpen
  \bibfield  {author} {\bibinfo {author} {\bibfnamefont {P.}~\bibnamefont
  {Dai}},\ }\href {\doibase 10.1103/RevModPhys.87.855} {\bibfield  {journal}
  {\bibinfo  {journal} {Rev. Mod. Phys.}\ }\textbf {\bibinfo {volume} {87}},\
  \bibinfo {pages} {855} (\bibinfo {year} {2015})}\BibitemShut {NoStop}%
\bibitem [{\citenamefont {Pratt}\ \emph {et~al.}(2009)\citenamefont {Pratt},
  \citenamefont {Tian}, \citenamefont {Kreyssig}, \citenamefont {Zarestky},
  \citenamefont {Nandi}, \citenamefont {Ni}, \citenamefont {Bud'ko},
  \citenamefont {Canfield}, \citenamefont {Goldman},\ and\ \citenamefont
  {McQueeney}}]{Pratt2009}%
  \BibitemOpen
  \bibfield  {author} {\bibinfo {author} {\bibfnamefont {D.~K.}\ \bibnamefont
  {Pratt}}, \bibinfo {author} {\bibfnamefont {W.}~\bibnamefont {Tian}},
  \bibinfo {author} {\bibfnamefont {A.}~\bibnamefont {Kreyssig}}, \bibinfo
  {author} {\bibfnamefont {J.~L.}\ \bibnamefont {Zarestky}}, \bibinfo {author}
  {\bibfnamefont {S.}~\bibnamefont {Nandi}}, \bibinfo {author} {\bibfnamefont
  {N.}~\bibnamefont {Ni}}, \bibinfo {author} {\bibfnamefont {S.~L.}\
  \bibnamefont {Bud'ko}}, \bibinfo {author} {\bibfnamefont {P.~C.}\
  \bibnamefont {Canfield}}, \bibinfo {author} {\bibfnamefont {A.~I.}\
  \bibnamefont {Goldman}}, \ and\ \bibinfo {author} {\bibfnamefont {R.~J.}\
  \bibnamefont {McQueeney}},\ }\href {\doibase 10.1103/PhysRevLett.103.087001}
  {\bibfield  {journal} {\bibinfo  {journal} {Phys. Rev. Lett.}\ }\textbf
  {\bibinfo {volume} {103}},\ \bibinfo {pages} {087001} (\bibinfo {year}
  {2009})}\BibitemShut {NoStop}%
\bibitem [{\citenamefont {Nandi}\ \emph {et~al.}(2010)\citenamefont {Nandi},
  \citenamefont {Kim}, \citenamefont {Kreyssig}, \citenamefont {Fernandes},
  \citenamefont {Pratt}, \citenamefont {Thaler}, \citenamefont {Ni},
  \citenamefont {Bud'ko}, \citenamefont {Canfield}, \citenamefont {Schmalian},
  \citenamefont {McQueeney},\ and\ \citenamefont {Goldman}}]{Nandi2010}%
  \BibitemOpen
  \bibfield  {author} {\bibinfo {author} {\bibfnamefont {S.}~\bibnamefont
  {Nandi}}, \bibinfo {author} {\bibfnamefont {M.~G.}\ \bibnamefont {Kim}},
  \bibinfo {author} {\bibfnamefont {A.}~\bibnamefont {Kreyssig}}, \bibinfo
  {author} {\bibfnamefont {R.~M.}\ \bibnamefont {Fernandes}}, \bibinfo {author}
  {\bibfnamefont {D.~K.}\ \bibnamefont {Pratt}}, \bibinfo {author}
  {\bibfnamefont {A.}~\bibnamefont {Thaler}}, \bibinfo {author} {\bibfnamefont
  {N.}~\bibnamefont {Ni}}, \bibinfo {author} {\bibfnamefont {S.~L.}\
  \bibnamefont {Bud'ko}}, \bibinfo {author} {\bibfnamefont {P.~C.}\
  \bibnamefont {Canfield}}, \bibinfo {author} {\bibfnamefont {J.}~\bibnamefont
  {Schmalian}}, \bibinfo {author} {\bibfnamefont {R.~J.}\ \bibnamefont
  {McQueeney}}, \ and\ \bibinfo {author} {\bibfnamefont {A.~I.}\ \bibnamefont
  {Goldman}},\ }\href {\doibase 10.1103/PhysRevLett.104.057006} {\bibfield
  {journal} {\bibinfo  {journal} {Phys. Rev. Lett.}\ }\textbf {\bibinfo
  {volume} {104}},\ \bibinfo {pages} {057006} (\bibinfo {year}
  {2010})}\BibitemShut {NoStop}%
\bibitem [{\citenamefont {Vollhardt}\ and\ \citenamefont
  {Wolfle}(2013)}]{vollhardt2013superfluid}%
  \BibitemOpen
  \bibfield  {author} {\bibinfo {author} {\bibfnamefont {D.}~\bibnamefont
  {Vollhardt}}\ and\ \bibinfo {author} {\bibfnamefont {P.}~\bibnamefont
  {Wolfle}},\ }\href@noop {} {\emph {\bibinfo {title} {The superfluid phases of
  helium 3}}}\ (\bibinfo  {publisher} {Courier Corporation},\ \bibinfo {year}
  {2013})\BibitemShut {NoStop}%
\bibitem [{\citenamefont {Balents}\ \emph {et~al.}(2020)\citenamefont
  {Balents}, \citenamefont {Dean}, \citenamefont {Efetov},\ and\ \citenamefont
  {Young}}]{balents2020superconductivity}%
  \BibitemOpen
  \bibfield  {author} {\bibinfo {author} {\bibfnamefont {L.}~\bibnamefont
  {Balents}}, \bibinfo {author} {\bibfnamefont {C.~R.}\ \bibnamefont {Dean}},
  \bibinfo {author} {\bibfnamefont {D.~K.}\ \bibnamefont {Efetov}}, \ and\
  \bibinfo {author} {\bibfnamefont {A.~F.}\ \bibnamefont {Young}},\ }\href
  {\doibase 10.1038/s41567-020-0906-9} {\bibfield  {journal} {\bibinfo
  {journal} {Nature Physics}\ }\textbf {\bibinfo {volume} {16}},\ \bibinfo
  {pages} {725} (\bibinfo {year} {2020})}\BibitemShut {NoStop}%
\bibitem [{\citenamefont {Anderson}(1987)}]{Anderson87}%
  \BibitemOpen
  \bibfield  {author} {\bibinfo {author} {\bibfnamefont {P.~W.}\ \bibnamefont
  {Anderson}},\ }\href {\doibase 10.1126/science.235.4793.1196} {\bibfield
  {journal} {\bibinfo  {journal} {Science}\ }\textbf {\bibinfo {volume}
  {235}},\ \bibinfo {pages} {1196} (\bibinfo {year} {1987})}\BibitemShut
  {NoStop}%
\bibitem [{\citenamefont {Wu}\ and\ \citenamefont {Zhang}(2005)}]{wuzhang2005}%
  \BibitemOpen
  \bibfield  {author} {\bibinfo {author} {\bibfnamefont {C.}~\bibnamefont
  {Wu}}\ and\ \bibinfo {author} {\bibfnamefont {S.-C.}\ \bibnamefont {Zhang}},\
  }\href {\doibase 10.1103/PhysRevB.71.155115} {\bibfield  {journal} {\bibinfo
  {journal} {Phys. Rev. B}\ }\textbf {\bibinfo {volume} {71}},\ \bibinfo
  {pages} {155115} (\bibinfo {year} {2005})}\BibitemShut {NoStop}%
\bibitem [{\citenamefont {Oshikawa}(2000)}]{Oshikawa00a}%
  \BibitemOpen
  \bibfield  {author} {\bibinfo {author} {\bibfnamefont {M.}~\bibnamefont
  {Oshikawa}},\ }\href {\doibase 10.1103/PhysRevLett.84.1535} {\bibfield
  {journal} {\bibinfo  {journal} {Phys. Rev. Lett.}\ }\textbf {\bibinfo
  {volume} {84}},\ \bibinfo {pages} {1535} (\bibinfo {year}
  {2000})}\BibitemShut {NoStop}%
\bibitem [{\citenamefont {Hastings}(2004)}]{Hasting04}%
  \BibitemOpen
  \bibfield  {author} {\bibinfo {author} {\bibfnamefont {M.~B.}\ \bibnamefont
  {Hastings}},\ }\href {\doibase 10.1103/PhysRevB.69.104431} {\bibfield
  {journal} {\bibinfo  {journal} {Phys. Rev. B}\ }\textbf {\bibinfo {volume}
  {69}},\ \bibinfo {pages} {104431} (\bibinfo {year} {2004})}\BibitemShut
  {NoStop}%
\bibitem [{\citenamefont {Lichtenstein}\ and\ \citenamefont
  {Katsnelson}(2000)}]{Lichtenstein2000}%
  \BibitemOpen
  \bibfield  {author} {\bibinfo {author} {\bibfnamefont {A.~I.}\ \bibnamefont
  {Lichtenstein}}\ and\ \bibinfo {author} {\bibfnamefont {M.~I.}\ \bibnamefont
  {Katsnelson}},\ }\href {\doibase 10.1103/PhysRevB.62.R9283} {\bibfield
  {journal} {\bibinfo  {journal} {Phys. Rev. B}\ }\textbf {\bibinfo {volume}
  {62}},\ \bibinfo {pages} {R9283} (\bibinfo {year} {2000})}\BibitemShut
  {NoStop}%
\bibitem [{\citenamefont {Jarrell}\ \emph {et~al.}(2001)\citenamefont
  {Jarrell}, \citenamefont {Maier}, \citenamefont {Hettler},\ and\
  \citenamefont {Tahvildarzadeh}}]{Jarrell2001}%
  \BibitemOpen
  \bibfield  {author} {\bibinfo {author} {\bibfnamefont {M.}~\bibnamefont
  {Jarrell}}, \bibinfo {author} {\bibfnamefont {T.}~\bibnamefont {Maier}},
  \bibinfo {author} {\bibfnamefont {M.~H.}\ \bibnamefont {Hettler}}, \ and\
  \bibinfo {author} {\bibfnamefont {A.~N.}\ \bibnamefont {Tahvildarzadeh}},\
  }\href {\doibase 10.1209/epl/i2001-00557-x} {\bibfield  {journal} {\bibinfo
  {journal} {Europhysics Letters ({EPL})}\ }\textbf {\bibinfo {volume} {56}},\
  \bibinfo {pages} {563} (\bibinfo {year} {2001})}\BibitemShut {NoStop}%
\bibitem [{\citenamefont {S\'en\'echal}\ \emph {et~al.}(2005)\citenamefont
  {S\'en\'echal}, \citenamefont {Lavertu}, \citenamefont {Marois},\ and\
  \citenamefont {Tremblay}}]{Senechal2005}%
  \BibitemOpen
  \bibfield  {author} {\bibinfo {author} {\bibfnamefont {D.}~\bibnamefont
  {S\'en\'echal}}, \bibinfo {author} {\bibfnamefont {P.-L.}\ \bibnamefont
  {Lavertu}}, \bibinfo {author} {\bibfnamefont {M.-A.}\ \bibnamefont {Marois}},
  \ and\ \bibinfo {author} {\bibfnamefont {A.-M.~S.}\ \bibnamefont
  {Tremblay}},\ }\href {\doibase 10.1103/PhysRevLett.94.156404} {\bibfield
  {journal} {\bibinfo  {journal} {Phys. Rev. Lett.}\ }\textbf {\bibinfo
  {volume} {94}},\ \bibinfo {pages} {156404} (\bibinfo {year}
  {2005})}\BibitemShut {NoStop}%
\bibitem [{\citenamefont {Capone}\ and\ \citenamefont
  {Kotliar}(2006)}]{Capone2006}%
  \BibitemOpen
  \bibfield  {author} {\bibinfo {author} {\bibfnamefont {M.}~\bibnamefont
  {Capone}}\ and\ \bibinfo {author} {\bibfnamefont {G.}~\bibnamefont
  {Kotliar}},\ }\href {\doibase 10.1103/PhysRevB.74.054513} {\bibfield
  {journal} {\bibinfo  {journal} {Phys. Rev. B}\ }\textbf {\bibinfo {volume}
  {74}},\ \bibinfo {pages} {054513} (\bibinfo {year} {2006})}\BibitemShut
  {NoStop}%
\bibitem [{\citenamefont {Nevidomskyy}\ \emph {et~al.}(2008)\citenamefont
  {Nevidomskyy}, \citenamefont {Scheiber}, \citenamefont {S\'en\'echal},\ and\
  \citenamefont {Tremblay}}]{Nevidomskyy2008}%
  \BibitemOpen
  \bibfield  {author} {\bibinfo {author} {\bibfnamefont {A.~H.}\ \bibnamefont
  {Nevidomskyy}}, \bibinfo {author} {\bibfnamefont {C.}~\bibnamefont
  {Scheiber}}, \bibinfo {author} {\bibfnamefont {D.}~\bibnamefont
  {S\'en\'echal}}, \ and\ \bibinfo {author} {\bibfnamefont {A.-M.~S.}\
  \bibnamefont {Tremblay}},\ }\href {\doibase 10.1103/PhysRevB.77.064427}
  {\bibfield  {journal} {\bibinfo  {journal} {Phys. Rev. B}\ }\textbf {\bibinfo
  {volume} {77}},\ \bibinfo {pages} {064427} (\bibinfo {year}
  {2008})}\BibitemShut {NoStop}%
\bibitem [{\citenamefont {Kancharla}\ \emph {et~al.}(2008)\citenamefont
  {Kancharla}, \citenamefont {Kyung}, \citenamefont {S\'en\'echal},
  \citenamefont {Civelli}, \citenamefont {Capone}, \citenamefont {Kotliar},\
  and\ \citenamefont {Tremblay}}]{Kancharla2008}%
  \BibitemOpen
  \bibfield  {author} {\bibinfo {author} {\bibfnamefont {S.~S.}\ \bibnamefont
  {Kancharla}}, \bibinfo {author} {\bibfnamefont {B.}~\bibnamefont {Kyung}},
  \bibinfo {author} {\bibfnamefont {D.}~\bibnamefont {S\'en\'echal}}, \bibinfo
  {author} {\bibfnamefont {M.}~\bibnamefont {Civelli}}, \bibinfo {author}
  {\bibfnamefont {M.}~\bibnamefont {Capone}}, \bibinfo {author} {\bibfnamefont
  {G.}~\bibnamefont {Kotliar}}, \ and\ \bibinfo {author} {\bibfnamefont
  {A.-M.~S.}\ \bibnamefont {Tremblay}},\ }\href {\doibase
  10.1103/PhysRevB.77.184516} {\bibfield  {journal} {\bibinfo  {journal} {Phys.
  Rev. B}\ }\textbf {\bibinfo {volume} {77}},\ \bibinfo {pages} {184516}
  (\bibinfo {year} {2008})}\BibitemShut {NoStop}%
\bibitem [{sup()}]{suppl}%
  \BibitemOpen
  \href@noop {} {\bibinfo  {journal} {See {\textrm{Supplemental Material}}(SM)
  for more details. The SM also contains additional
  Refs.~\cite{Assaad2005,Van1996,Sandvik1998}.}\ }\BibitemShut {NoStop}%
\bibitem [{\citenamefont {Blankenbecler}\ \emph {et~al.}(1981)\citenamefont
  {Blankenbecler}, \citenamefont {Scalapino},\ and\ \citenamefont
  {Sugar}}]{BSS}%
  \BibitemOpen
\bibfield  {journal} {  }\bibfield  {author} {\bibinfo {author} {\bibfnamefont
  {R.}~\bibnamefont {Blankenbecler}}, \bibinfo {author} {\bibfnamefont {D.~J.}\
  \bibnamefont {Scalapino}}, \ and\ \bibinfo {author} {\bibfnamefont {R.~L.}\
  \bibnamefont {Sugar}},\ }\href {\doibase 10.1103/PhysRevD.24.2278} {\bibfield
   {journal} {\bibinfo  {journal} {Phys. Rev. D}\ }\textbf {\bibinfo {volume}
  {24}},\ \bibinfo {pages} {2278} (\bibinfo {year} {1981})}\BibitemShut
  {NoStop}%
\bibitem [{\citenamefont {Assaad}\ and\ \citenamefont
  {Evertz}(2008)}]{AssaadEvertz2008}%
  \BibitemOpen
  \bibfield  {author} {\bibinfo {author} {\bibfnamefont {F.}~\bibnamefont
  {Assaad}}\ and\ \bibinfo {author} {\bibfnamefont {H.}~\bibnamefont
  {Evertz}},\ }in\ \href {\doibase 10.1007/978-3-540-74686-7_10} {\emph
  {\bibinfo {booktitle} {Computational Many-Particle Physics}}},\ \bibinfo
  {series} {Lecture Notes in Physics}, Vol.\ \bibinfo {volume} {739},\ \bibinfo
  {editor} {edited by\ \bibinfo {editor} {\bibfnamefont {H.}~\bibnamefont
  {Fehske}}, \bibinfo {editor} {\bibfnamefont {R.}~\bibnamefont {Schneider}}, \
  and\ \bibinfo {editor} {\bibfnamefont {A.}~\bibnamefont {Wei{\ss}e}}}\
  (\bibinfo  {publisher} {Springer Berlin Heidelberg},\ \bibinfo {year}
  {2008})\ pp.\ \bibinfo {pages} {277--356}\BibitemShut {NoStop}%
\bibitem [{\citenamefont {Hirsch}(1983)}]{Hirsch83}%
  \BibitemOpen
  \bibfield  {author} {\bibinfo {author} {\bibfnamefont {J.}~\bibnamefont
  {Hirsch}},\ }\href {\doibase 10.1103/PhysRevB.28.4059} {\bibfield  {journal}
  {\bibinfo  {journal} {Phys. Rev. B}\ }\textbf {\bibinfo {volume} {28}},\
  \bibinfo {pages} {4059} (\bibinfo {year} {1983})}\BibitemShut {NoStop}%
\bibitem [{\citenamefont {Balents}\ \emph {et~al.}(1998)\citenamefont
  {Balents}, \citenamefont {Fisher},\ and\ \citenamefont
  {Nayak}}]{balents1998nodal}%
  \BibitemOpen
  \bibfield  {author} {\bibinfo {author} {\bibfnamefont {L.}~\bibnamefont
  {Balents}}, \bibinfo {author} {\bibfnamefont {M.~P.~A.}\ \bibnamefont
  {Fisher}}, \ and\ \bibinfo {author} {\bibfnamefont {C.}~\bibnamefont
  {Nayak}},\ }\href {\doibase 10.1142/S0217979298000570} {\bibfield  {journal}
  {\bibinfo  {journal} {International Journal of Modern Physics B}\ }\textbf
  {\bibinfo {volume} {12}},\ \bibinfo {pages} {1033} (\bibinfo {year}
  {1998})}\BibitemShut {NoStop}%
\bibitem [{\citenamefont {Rosenstein}\ \emph {et~al.}(1993)\citenamefont
  {Rosenstein}, \citenamefont {{Hoi-Lai Yu}},\ and\ \citenamefont
  {Kovner}}]{Rosenstein1993}%
  \BibitemOpen
  \bibfield  {author} {\bibinfo {author} {\bibfnamefont {B.}~\bibnamefont
  {Rosenstein}}, \bibinfo {author} {\bibnamefont {{Hoi-Lai Yu}}}, \ and\
  \bibinfo {author} {\bibfnamefont {A.}~\bibnamefont {Kovner}},\ }\href
  {\doibase https://doi.org/10.1016/0370-2693(93)91253-J} {\bibfield  {journal}
  {\bibinfo  {journal} {Physics Letters B}\ }\textbf {\bibinfo {volume}
  {314}},\ \bibinfo {pages} {381 } (\bibinfo {year} {1993})}\BibitemShut
  {NoStop}%
\bibitem [{\citenamefont {Herbut}(2006)}]{Herbut06}%
  \BibitemOpen
  \bibfield  {author} {\bibinfo {author} {\bibfnamefont {I.~F.}\ \bibnamefont
  {Herbut}},\ }\href {\doibase 10.1103/PhysRevLett.97.146401} {\bibfield
  {journal} {\bibinfo  {journal} {Phys. Rev. Lett.}\ }\textbf {\bibinfo
  {volume} {97}},\ \bibinfo {pages} {146401} (\bibinfo {year}
  {2006})}\BibitemShut {NoStop}%
\bibitem [{\citenamefont {Herbut}\ \emph {et~al.}(2009)\citenamefont {Herbut},
  \citenamefont {Juri\ifmmode \check{c}\else \v{c}\fi{}i\ifmmode~\acute{c}\else
  \'{c}\fi{}},\ and\ \citenamefont {Roy}}]{Herbut09}%
  \BibitemOpen
  \bibfield  {author} {\bibinfo {author} {\bibfnamefont {I.~F.}\ \bibnamefont
  {Herbut}}, \bibinfo {author} {\bibfnamefont {V.}~\bibnamefont {Juri\ifmmode
  \check{c}\else \v{c}\fi{}i\ifmmode~\acute{c}\else \'{c}\fi{}}}, \ and\
  \bibinfo {author} {\bibfnamefont {B.}~\bibnamefont {Roy}},\ }\href {\doibase
  10.1103/PhysRevB.79.085116} {\bibfield  {journal} {\bibinfo  {journal} {Phys.
  Rev. B}\ }\textbf {\bibinfo {volume} {79}},\ \bibinfo {pages} {085116}
  (\bibinfo {year} {2009})}\BibitemShut {NoStop}%
\bibitem [{\citenamefont {Assaad}\ and\ \citenamefont
  {Herbut}(2013)}]{Assaad13}%
  \BibitemOpen
  \bibfield  {author} {\bibinfo {author} {\bibfnamefont {F.~F.}\ \bibnamefont
  {Assaad}}\ and\ \bibinfo {author} {\bibfnamefont {I.~F.}\ \bibnamefont
  {Herbut}},\ }\href {\doibase 10.1103/PhysRevX.3.031010} {\bibfield  {journal}
  {\bibinfo  {journal} {Phys. Rev. X}\ }\textbf {\bibinfo {volume} {3}},\
  \bibinfo {pages} {031010} (\bibinfo {year} {2013})}\BibitemShut {NoStop}%
\bibitem [{\citenamefont {Janssen}\ and\ \citenamefont
  {Herbut}(2014)}]{Janssen2014}%
  \BibitemOpen
  \bibfield  {author} {\bibinfo {author} {\bibfnamefont {L.}~\bibnamefont
  {Janssen}}\ and\ \bibinfo {author} {\bibfnamefont {I.~F.}\ \bibnamefont
  {Herbut}},\ }\href {\doibase 10.1103/PhysRevB.89.205403} {\bibfield
  {journal} {\bibinfo  {journal} {Phys. Rev. B}\ }\textbf {\bibinfo {volume}
  {89}},\ \bibinfo {pages} {205403} (\bibinfo {year} {2014})}\BibitemShut
  {NoStop}%
\bibitem [{\citenamefont {Parisen~Toldin}\ \emph {et~al.}(2015)\citenamefont
  {Parisen~Toldin}, \citenamefont {Hohenadler}, \citenamefont {Assaad},\ and\
  \citenamefont {Herbut}}]{Toldin14}%
  \BibitemOpen
  \bibfield  {author} {\bibinfo {author} {\bibfnamefont {F.}~\bibnamefont
  {Parisen~Toldin}}, \bibinfo {author} {\bibfnamefont {M.}~\bibnamefont
  {Hohenadler}}, \bibinfo {author} {\bibfnamefont {F.~F.}\ \bibnamefont
  {Assaad}}, \ and\ \bibinfo {author} {\bibfnamefont {I.~F.}\ \bibnamefont
  {Herbut}},\ }\href {\doibase 10.1103/PhysRevB.91.165108} {\bibfield
  {journal} {\bibinfo  {journal} {Phys. Rev. B}\ }\textbf {\bibinfo {volume}
  {91}},\ \bibinfo {pages} {165108} (\bibinfo {year} {2015})}\BibitemShut
  {NoStop}%
\bibitem [{\citenamefont {Otsuka}\ \emph {et~al.}(2016)\citenamefont {Otsuka},
  \citenamefont {Yunoki},\ and\ \citenamefont {Sorella}}]{Otsuka16}%
  \BibitemOpen
  \bibfield  {author} {\bibinfo {author} {\bibfnamefont {Y.}~\bibnamefont
  {Otsuka}}, \bibinfo {author} {\bibfnamefont {S.}~\bibnamefont {Yunoki}}, \
  and\ \bibinfo {author} {\bibfnamefont {S.}~\bibnamefont {Sorella}},\ }\href
  {\doibase 10.1103/PhysRevX.6.011029} {\bibfield  {journal} {\bibinfo
  {journal} {Phys. Rev. X}\ }\textbf {\bibinfo {volume} {6}},\ \bibinfo {pages}
  {011029} (\bibinfo {year} {2016})}\BibitemShut {NoStop}%
\bibitem [{\citenamefont {Zerf}\ \emph {et~al.}(2017)\citenamefont {Zerf},
  \citenamefont {Mihaila}, \citenamefont {Marquard}, \citenamefont {Herbut},\
  and\ \citenamefont {Scherer}}]{Zerf2017}%
  \BibitemOpen
  \bibfield  {author} {\bibinfo {author} {\bibfnamefont {N.}~\bibnamefont
  {Zerf}}, \bibinfo {author} {\bibfnamefont {L.~N.}\ \bibnamefont {Mihaila}},
  \bibinfo {author} {\bibfnamefont {P.}~\bibnamefont {Marquard}}, \bibinfo
  {author} {\bibfnamefont {I.~F.}\ \bibnamefont {Herbut}}, \ and\ \bibinfo
  {author} {\bibfnamefont {M.~M.}\ \bibnamefont {Scherer}},\ }\href {\doibase
  10.1103/PhysRevD.96.096010} {\bibfield  {journal} {\bibinfo  {journal} {Phys.
  Rev. D}\ }\textbf {\bibinfo {volume} {96}},\ \bibinfo {pages} {096010}
  (\bibinfo {year} {2017})}\BibitemShut {NoStop}%
\bibitem [{\citenamefont {Gracey}(2018)}]{Gracey2018}%
  \BibitemOpen
  \bibfield  {author} {\bibinfo {author} {\bibfnamefont {J.~A.}\ \bibnamefont
  {Gracey}},\ }\href {\doibase 10.1103/PhysRevD.97.105009} {\bibfield
  {journal} {\bibinfo  {journal} {Phys. Rev. D}\ }\textbf {\bibinfo {volume}
  {97}},\ \bibinfo {pages} {105009} (\bibinfo {year} {2018})}\BibitemShut
  {NoStop}%
\bibitem [{\citenamefont {Buividovich}\ \emph {et~al.}(2018)\citenamefont
  {Buividovich}, \citenamefont {Smith}, \citenamefont {Ulybyshev},\ and\
  \citenamefont {von Smekal}}]{Buividovich2018}%
  \BibitemOpen
  \bibfield  {author} {\bibinfo {author} {\bibfnamefont {P.}~\bibnamefont
  {Buividovich}}, \bibinfo {author} {\bibfnamefont {D.}~\bibnamefont {Smith}},
  \bibinfo {author} {\bibfnamefont {M.}~\bibnamefont {Ulybyshev}}, \ and\
  \bibinfo {author} {\bibfnamefont {L.}~\bibnamefont {von Smekal}},\ }\href
  {\doibase 10.1103/PhysRevB.98.235129} {\bibfield  {journal} {\bibinfo
  {journal} {Phys. Rev. B}\ }\textbf {\bibinfo {volume} {98}},\ \bibinfo
  {pages} {235129} (\bibinfo {year} {2018})}\BibitemShut {NoStop}%
\bibitem [{\citenamefont {Lang}\ and\ \citenamefont
  {L\"auchli}(2019)}]{Lang2019}%
  \BibitemOpen
  \bibfield  {author} {\bibinfo {author} {\bibfnamefont {T.~C.}\ \bibnamefont
  {Lang}}\ and\ \bibinfo {author} {\bibfnamefont {A.~M.}\ \bibnamefont
  {L\"auchli}},\ }\href {\doibase 10.1103/PhysRevLett.123.137602} {\bibfield
  {journal} {\bibinfo  {journal} {Phys. Rev. Lett.}\ }\textbf {\bibinfo
  {volume} {123}},\ \bibinfo {pages} {137602} (\bibinfo {year}
  {2019})}\BibitemShut {NoStop}%
\bibitem [{\citenamefont {Otsuka}\ \emph {et~al.}(2020)\citenamefont {Otsuka},
  \citenamefont {Seki}, \citenamefont {Sorella},\ and\ \citenamefont
  {Yunoki}}]{otsuka2020dirac}%
  \BibitemOpen
  \bibfield  {author} {\bibinfo {author} {\bibfnamefont {Y.}~\bibnamefont
  {Otsuka}}, \bibinfo {author} {\bibfnamefont {K.}~\bibnamefont {Seki}},
  \bibinfo {author} {\bibfnamefont {S.}~\bibnamefont {Sorella}}, \ and\
  \bibinfo {author} {\bibfnamefont {S.}~\bibnamefont {Yunoki}},\ }\href@noop {}
  {\bibfield  {journal} {\bibinfo  {journal} {arXiv:2009.04685
  [cond-mat.str-el]}\ } (\bibinfo {year} {2020})}\BibitemShut {NoStop}%
\bibitem [{\citenamefont {Berg}\ \emph {et~al.}(2012)\citenamefont {Berg},
  \citenamefont {Metlitski},\ and\ \citenamefont {Sachdev}}]{Berg12}%
  \BibitemOpen
  \bibfield  {author} {\bibinfo {author} {\bibfnamefont {E.}~\bibnamefont
  {Berg}}, \bibinfo {author} {\bibfnamefont {M.~A.}\ \bibnamefont {Metlitski}},
  \ and\ \bibinfo {author} {\bibfnamefont {S.}~\bibnamefont {Sachdev}},\ }\href
  {\doibase 10.1126/science.1227769} {\bibfield  {journal} {\bibinfo  {journal}
  {Science}\ }\textbf {\bibinfo {volume} {338}},\ \bibinfo {pages} {1606}
  (\bibinfo {year} {2012})}\BibitemShut {NoStop}%
\bibitem [{\citenamefont {Schattner}\ \emph
  {et~al.}(2016{\natexlab{a}})\citenamefont {Schattner}, \citenamefont
  {Lederer}, \citenamefont {Kivelson},\ and\ \citenamefont
  {Berg}}]{Schattner2016}%
  \BibitemOpen
  \bibfield  {author} {\bibinfo {author} {\bibfnamefont {Y.}~\bibnamefont
  {Schattner}}, \bibinfo {author} {\bibfnamefont {S.}~\bibnamefont {Lederer}},
  \bibinfo {author} {\bibfnamefont {S.~A.}\ \bibnamefont {Kivelson}}, \ and\
  \bibinfo {author} {\bibfnamefont {E.}~\bibnamefont {Berg}},\ }\href {\doibase
  10.1103/PhysRevX.6.031028} {\bibfield  {journal} {\bibinfo  {journal} {Phys.
  Rev. X}\ }\textbf {\bibinfo {volume} {6}},\ \bibinfo {pages} {031028}
  (\bibinfo {year} {2016}{\natexlab{a}})}\BibitemShut {NoStop}%
\bibitem [{\citenamefont {Berg}\ \emph {et~al.}(2019)\citenamefont {Berg},
  \citenamefont {Lederer}, \citenamefont {Schattner},\ and\ \citenamefont
  {Trebst}}]{Berg2019}%
  \BibitemOpen
  \bibfield  {author} {\bibinfo {author} {\bibfnamefont {E.}~\bibnamefont
  {Berg}}, \bibinfo {author} {\bibfnamefont {S.}~\bibnamefont {Lederer}},
  \bibinfo {author} {\bibfnamefont {Y.}~\bibnamefont {Schattner}}, \ and\
  \bibinfo {author} {\bibfnamefont {S.}~\bibnamefont {Trebst}},\ }\href
  {\doibase 10.1146/annurev-conmatphys-031218-013339} {\bibfield  {journal}
  {\bibinfo  {journal} {Annual Review of Condensed Matter Physics}\ }\textbf
  {\bibinfo {volume} {10}},\ \bibinfo {pages} {63} (\bibinfo {year}
  {2019})}\BibitemShut {NoStop}%
\bibitem [{\citenamefont {Xu}\ \emph {et~al.}(2019)\citenamefont {Xu},
  \citenamefont {Liu}, \citenamefont {Pan}, \citenamefont {Qi}, \citenamefont
  {Sun},\ and\ \citenamefont {Meng}}]{Xu2019}%
  \BibitemOpen
  \bibfield  {author} {\bibinfo {author} {\bibfnamefont {X.~Y.}\ \bibnamefont
  {Xu}}, \bibinfo {author} {\bibfnamefont {Z.~H.}\ \bibnamefont {Liu}},
  \bibinfo {author} {\bibfnamefont {G.}~\bibnamefont {Pan}}, \bibinfo {author}
  {\bibfnamefont {Y.}~\bibnamefont {Qi}}, \bibinfo {author} {\bibfnamefont
  {K.}~\bibnamefont {Sun}}, \ and\ \bibinfo {author} {\bibfnamefont {Z.~Y.}\
  \bibnamefont {Meng}},\ }\href {\doibase 10.1088/1361-648x/ab3295} {\bibfield
  {journal} {\bibinfo  {journal} {Journal of Physics: Condensed Matter}\
  }\textbf {\bibinfo {volume} {31}},\ \bibinfo {pages} {463001} (\bibinfo
  {year} {2019})}\BibitemShut {NoStop}%
\bibitem [{\citenamefont {Li}\ and\ \citenamefont {Yao}(2019)}]{Li2019}%
  \BibitemOpen
  \bibfield  {author} {\bibinfo {author} {\bibfnamefont {Z.-X.}\ \bibnamefont
  {Li}}\ and\ \bibinfo {author} {\bibfnamefont {H.}~\bibnamefont {Yao}},\
  }\href {\doibase 10.1146/annurev-conmatphys-033117-054307} {\bibfield
  {journal} {\bibinfo  {journal} {Annual Review of Condensed Matter Physics}\
  }\textbf {\bibinfo {volume} {10}},\ \bibinfo {pages} {337} (\bibinfo {year}
  {2019})}\BibitemShut {NoStop}%
\bibitem [{\citenamefont {Schattner}\ \emph
  {et~al.}(2016{\natexlab{b}})\citenamefont {Schattner}, \citenamefont
  {Gerlach}, \citenamefont {Trebst},\ and\ \citenamefont {Berg}}]{Berg2016}%
  \BibitemOpen
  \bibfield  {author} {\bibinfo {author} {\bibfnamefont {Y.}~\bibnamefont
  {Schattner}}, \bibinfo {author} {\bibfnamefont {M.~H.}\ \bibnamefont
  {Gerlach}}, \bibinfo {author} {\bibfnamefont {S.}~\bibnamefont {Trebst}}, \
  and\ \bibinfo {author} {\bibfnamefont {E.}~\bibnamefont {Berg}},\ }\href
  {\doibase 10.1103/PhysRevLett.117.097002} {\bibfield  {journal} {\bibinfo
  {journal} {Phys. Rev. Lett.}\ }\textbf {\bibinfo {volume} {117}},\ \bibinfo
  {pages} {097002} (\bibinfo {year} {2016}{\natexlab{b}})}\BibitemShut
  {NoStop}%
\bibitem [{\citenamefont {Dumitrescu}\ \emph {et~al.}(2016)\citenamefont
  {Dumitrescu}, \citenamefont {Serbyn}, \citenamefont {Scalettar},\ and\
  \citenamefont {Vishwanath}}]{dumitrescu2016}%
  \BibitemOpen
  \bibfield  {author} {\bibinfo {author} {\bibfnamefont {P.~T.}\ \bibnamefont
  {Dumitrescu}}, \bibinfo {author} {\bibfnamefont {M.}~\bibnamefont {Serbyn}},
  \bibinfo {author} {\bibfnamefont {R.~T.}\ \bibnamefont {Scalettar}}, \ and\
  \bibinfo {author} {\bibfnamefont {A.}~\bibnamefont {Vishwanath}},\ }\href
  {\doibase 10.1103/PhysRevB.94.155127} {\bibfield  {journal} {\bibinfo
  {journal} {Phys. Rev. B}\ }\textbf {\bibinfo {volume} {94}},\ \bibinfo
  {pages} {155127} (\bibinfo {year} {2016})}\BibitemShut {NoStop}%
\bibitem [{\citenamefont {Li}\ \emph {et~al.}(2017)\citenamefont {Li},
  \citenamefont {Wang}, \citenamefont {Yao},\ and\ \citenamefont
  {Lee}}]{li2017nature}%
  \BibitemOpen
  \bibfield  {author} {\bibinfo {author} {\bibfnamefont {Z.-X.}\ \bibnamefont
  {Li}}, \bibinfo {author} {\bibfnamefont {F.}~\bibnamefont {Wang}}, \bibinfo
  {author} {\bibfnamefont {H.}~\bibnamefont {Yao}}, \ and\ \bibinfo {author}
  {\bibfnamefont {D.-H.}\ \bibnamefont {Lee}},\ }\href {\doibase
  10.1103/PhysRevB.95.214505} {\bibfield  {journal} {\bibinfo  {journal} {Phys.
  Rev. B}\ }\textbf {\bibinfo {volume} {95}},\ \bibinfo {pages} {214505}
  (\bibinfo {year} {2017})}\BibitemShut {NoStop}%
\bibitem [{\citenamefont {Lederer}\ \emph {et~al.}(2017)\citenamefont
  {Lederer}, \citenamefont {Schattner}, \citenamefont {Berg},\ and\
  \citenamefont {Kivelson}}]{lederer2017superconductivity}%
  \BibitemOpen
  \bibfield  {author} {\bibinfo {author} {\bibfnamefont {S.}~\bibnamefont
  {Lederer}}, \bibinfo {author} {\bibfnamefont {Y.}~\bibnamefont {Schattner}},
  \bibinfo {author} {\bibfnamefont {E.}~\bibnamefont {Berg}}, \ and\ \bibinfo
  {author} {\bibfnamefont {S.~A.}\ \bibnamefont {Kivelson}},\ }\href {\doibase
  10.1073/pnas.1620651114} {\bibfield  {journal} {\bibinfo  {journal}
  {Proceedings of the National Academy of Sciences}\ }\textbf {\bibinfo
  {volume} {114}},\ \bibinfo {pages} {4905} (\bibinfo {year}
  {2017})}\BibitemShut {NoStop}%
\bibitem [{\citenamefont {Wang}\ \emph {et~al.}(2017)\citenamefont {Wang},
  \citenamefont {Schattner}, \citenamefont {Berg},\ and\ \citenamefont
  {Fernandes}}]{Wang2017}%
  \BibitemOpen
  \bibfield  {author} {\bibinfo {author} {\bibfnamefont {X.}~\bibnamefont
  {Wang}}, \bibinfo {author} {\bibfnamefont {Y.}~\bibnamefont {Schattner}},
  \bibinfo {author} {\bibfnamefont {E.}~\bibnamefont {Berg}}, \ and\ \bibinfo
  {author} {\bibfnamefont {R.~M.}\ \bibnamefont {Fernandes}},\ }\href {\doibase
  10.1103/PhysRevB.95.174520} {\bibfield  {journal} {\bibinfo  {journal} {Phys.
  Rev. B}\ }\textbf {\bibinfo {volume} {95}},\ \bibinfo {pages} {174520}
  (\bibinfo {year} {2017})}\BibitemShut {NoStop}%
\bibitem [{\citenamefont {Christensen}\ \emph {et~al.}(2020)\citenamefont
  {Christensen}, \citenamefont {Wang}, \citenamefont {Schattner}, \citenamefont
  {Berg},\ and\ \citenamefont {Fernandes}}]{christensen2020modeling}%
  \BibitemOpen
  \bibfield  {author} {\bibinfo {author} {\bibfnamefont {M.~H.}\ \bibnamefont
  {Christensen}}, \bibinfo {author} {\bibfnamefont {X.}~\bibnamefont {Wang}},
  \bibinfo {author} {\bibfnamefont {Y.}~\bibnamefont {Schattner}}, \bibinfo
  {author} {\bibfnamefont {E.}~\bibnamefont {Berg}}, \ and\ \bibinfo {author}
  {\bibfnamefont {R.~M.}\ \bibnamefont {Fernandes}},\ }\href@noop {} {\bibfield
   {journal} {\bibinfo  {journal} {arXiv:2006.11203 [cond-mat.str-el]}\ }
  (\bibinfo {year} {2020})}\BibitemShut {NoStop}%
\bibitem [{\citenamefont {Assaad}\ \emph {et~al.}(1996)\citenamefont {Assaad},
  \citenamefont {Imada},\ and\ \citenamefont {Scalapino}}]{assaad1996quantum}%
  \BibitemOpen
  \bibfield  {author} {\bibinfo {author} {\bibfnamefont {F.~F.}\ \bibnamefont
  {Assaad}}, \bibinfo {author} {\bibfnamefont {M.}~\bibnamefont {Imada}}, \
  and\ \bibinfo {author} {\bibfnamefont {D.~J.}\ \bibnamefont {Scalapino}},\
  }\href {\doibase 10.1103/PhysRevLett.77.4592} {\bibfield  {journal} {\bibinfo
   {journal} {Phys. Rev. Lett.}\ }\textbf {\bibinfo {volume} {77}},\ \bibinfo
  {pages} {4592} (\bibinfo {year} {1996})}\BibitemShut {NoStop}%
\bibitem [{\citenamefont {Towns}\ \emph {et~al.}(2014)\citenamefont {Towns},
  \citenamefont {Cockerill}, \citenamefont {Dahan}, \citenamefont {Foster},
  \citenamefont {Gaither}, \citenamefont {Grimshaw}, \citenamefont {Hazlewood},
  \citenamefont {Lathrop}, \citenamefont {Lifka}, \citenamefont {Peterson},
  \citenamefont {Roskies}, \citenamefont {Scott},\ and\ \citenamefont
  {Wilkins-Diehr}}]{xsede}%
  \BibitemOpen
  \bibfield  {author} {\bibinfo {author} {\bibfnamefont {J.}~\bibnamefont
  {Towns}}, \bibinfo {author} {\bibfnamefont {T.}~\bibnamefont {Cockerill}},
  \bibinfo {author} {\bibfnamefont {M.}~\bibnamefont {Dahan}}, \bibinfo
  {author} {\bibfnamefont {I.}~\bibnamefont {Foster}}, \bibinfo {author}
  {\bibfnamefont {K.}~\bibnamefont {Gaither}}, \bibinfo {author} {\bibfnamefont
  {A.}~\bibnamefont {Grimshaw}}, \bibinfo {author} {\bibfnamefont
  {V.}~\bibnamefont {Hazlewood}}, \bibinfo {author} {\bibfnamefont
  {S.}~\bibnamefont {Lathrop}}, \bibinfo {author} {\bibfnamefont
  {D.}~\bibnamefont {Lifka}}, \bibinfo {author} {\bibfnamefont {G.~D.}\
  \bibnamefont {Peterson}}, \bibinfo {author} {\bibfnamefont {R.}~\bibnamefont
  {Roskies}}, \bibinfo {author} {\bibfnamefont {J.}~\bibnamefont {Scott}}, \
  and\ \bibinfo {author} {\bibfnamefont {N.}~\bibnamefont {Wilkins-Diehr}},\
  }\href {\doibase 10.1109/MCSE.2014.80} {\bibfield  {journal} {\bibinfo
  {journal} {Computing in Science and Engineering}\ }\textbf {\bibinfo {volume}
  {16}},\ \bibinfo {pages} {62} (\bibinfo {year} {2014})}\BibitemShut {NoStop}%
\bibitem [{\citenamefont {Assaad}(2005)}]{Assaad2005}%
  \BibitemOpen
  \bibfield  {author} {\bibinfo {author} {\bibfnamefont {F.~F.}\ \bibnamefont
  {Assaad}},\ }\href {\doibase 10.1103/PhysRevB.71.075103} {\bibfield
  {journal} {\bibinfo  {journal} {Phys. Rev. B}\ }\textbf {\bibinfo {volume}
  {71}},\ \bibinfo {pages} {075103} (\bibinfo {year} {2005})}\BibitemShut
  {NoStop}%
\bibitem [{\citenamefont {van Leeuwen}\ \emph {et~al.}(1996)\citenamefont {van
  Leeuwen}, \citenamefont {du~Croo~de Jongh},\ and\ \citenamefont
  {Denteneer}}]{Van1996}%
  \BibitemOpen
  \bibfield  {author} {\bibinfo {author} {\bibfnamefont {J.~M.~J.}\
  \bibnamefont {van Leeuwen}}, \bibinfo {author} {\bibfnamefont {M.~S.~L.}\
  \bibnamefont {du~Croo~de Jongh}}, \ and\ \bibinfo {author} {\bibfnamefont
  {P.~J.~H.}\ \bibnamefont {Denteneer}},\ }\href {\doibase
  10.1088/0305-4470/29/1/008} {\bibfield  {journal} {\bibinfo  {journal}
  {Journal of Physics A: Mathematical and General}\ }\textbf {\bibinfo {volume}
  {29}},\ \bibinfo {pages} {41} (\bibinfo {year} {1996})}\BibitemShut {NoStop}%
\bibitem [{\citenamefont {Sandvik}(1998)}]{Sandvik1998}%
  \BibitemOpen
  \bibfield  {author} {\bibinfo {author} {\bibfnamefont {A.~W.}\ \bibnamefont
  {Sandvik}},\ }\href {\doibase 10.1103/PhysRevB.57.10287} {\bibfield
  {journal} {\bibinfo  {journal} {Phys. Rev. B}\ }\textbf {\bibinfo {volume}
  {57}},\ \bibinfo {pages} {10287} (\bibinfo {year} {1998})}\BibitemShut
  {NoStop}%
\end{thebibliography}%

\newpage
\clearpage
\onecolumngrid
\begin{center}
\textbf{Supplementary Material for ``Competing nodal d-wave superconductivity and antiferromagnetism"}
\end{center}
\setcounter{equation}{0}
\setcounter{figure}{0}
\setcounter{table}{0}
\setcounter{page}{1}
\makeatletter
\renewcommand{\thetable}{S\arabic{table}}
\renewcommand{\theequation}{S\arabic{equation}}
\renewcommand{\thefigure}{S\arabic{figure}}
\setcounter{secnumdepth}{3}

\onecolumngrid
\setcounter{equation}{0}
\setcounter{figure}{0}
\setcounter{table}{0}
\setcounter{page}{1}
\makeatletter
\renewcommand{\thetable}{S\arabic{table}}
\renewcommand{\theequation}{S\arabic{equation}}
\renewcommand{\thefigure}{S\arabic{figure}}
\renewcommand{\bibnumfmt}[1]{[S#1]}
\renewcommand{\citenumfont}[1]{S#1}
\setcounter{secnumdepth}{3}

\section{Mean-field theory}
As discussed in the main text,  the Hamiltonian of our model is:
$H = H_t + H_U + H_V + H_{\XY}$ with
\begin{align}
& H_{t}= -t\sum_{\langle ij\rangle,\sigma}(c_{i,\sigma}^{\dagger}c_{j,\sigma}+\text{h.c.}) \\
& H_{U}= \frac{U}{2}\sum_{i}({\rho}_{i,\uparrow}+{\rho}_{i,\downarrow}-1)^2 \\
& H_{V}= V \sum_{\langle ij\rangle}(\tau_{i,j}e^{i\theta_{i,j}}(c_{i,\uparrow}^{\dagger}c_{j,\downarrow}^{\dagger}-c_{j,\uparrow}^{\dagger}c_{i,\downarrow}^{\dagger})+\text{h.c.}) \\
& H_{\XY}= K\sum_{\langle ij\rangle}n_{ij}^{2}-J\sum_{\langle ij,kl\rangle}\cos(\theta_{ij}-\theta_{kl}),
\end{align}
Let's define 
$\hat{\alpha}_{ij}\equiv -e^{i\theta_{ij}}$, 
$\hat{\Delta}_{ij}\equiv \tau_{i,j} ( c_{i,\uparrow}^\dagger c_{j,\downarrow}^\dagger - c_{i,\downarrow}^\dagger c_{j,\uparrow}^\dagger ) $.
We consider mean-field order parameters $\alpha = \langle \hat{\alpha}_{ij} \rangle $, $\Delta = \langle \hat{\Delta}_{ij} \rangle$ and $m = (-)^{i}\langle (\rho_{i,\uparrow}-\rho_{i,\downarrow})\rangle$, leading to two coupled mean-field Hamiltonians. For fermions $c$ we have,
\begin{align}
H_\text{f}^{\text{MF}}= & -t\sum_{\langle ij \rangle,\sigma}\left(c_{i,\sigma}^{\dagger}c_{j,\sigma}+h.c.\right)-Um\sum_{i}(-)^{i}(\rho_{i,\uparrow}-\rho_{i,\downarrow}) \nonumber \\
 & - V\alpha\sum_{\langle ij \rangle} \left( \hat{\Delta}_{ij} + \text{h.c.} \right)
\end{align}
For rotors $\theta$ we have,
\begin{align}
H_{\theta}^{\text{MF}}= & K\sum_{\langle ij\rangle}n_{ij}^{2}-J\sum_{\langle ij,kl\rangle}\cos(\theta_{ij}-\theta_{kl}) \nonumber \\
 & - V\Delta\sum_{\langle ij \rangle} \left( \hat{\alpha}_{ij}+\text{h.c.} \right) \label{eq:rotormf}
\end{align}
We solve the two coupled mean-field Hamiltonians self-consistently.

$H_{\theta}^{\text{MF}}$ part can be solved with numerical exact diagonalization (ED). We consider $\theta$ fields living on only four bonds of a square plaquette, and solve the four-bond-problem with ED. Choosing an eigenbasis
of $n_{\delta}$, $n_{\delta}|l\rangle=l|l\rangle$, we truncate $|l|\le l_{c}$. (The results converge rather fast with $l_c$ and $l_c=4$ turned out to be sufficient). We first solve the following eigenvalue problem
\begin{equation}
H_{\theta}^{\text{MF}}|\psi_i(\theta)\rangle=E_i|\psi_i(\theta)\rangle,
\end{equation}
and obtain a self-consistent equation for $\alpha$:
\begin{equation}
\alpha=\frac{\sum_i\langle\psi_i(\theta)|\hat{\alpha}|\psi_i(\theta)\rangle e^{-\beta E_i}}{\sum_i e^{-\beta E_i}}.
\end{equation}

The fermion mean-field Hamiltonian $H_\text{f}^{\text{MF}}$ is quadratic, and can be diagonalized in a straightforward manner. This leads to self-consistent equations for $m$ and $\Delta$:
\begin{align}
 & m = \frac{U}{N}\sum_k \frac{m}{E(\vec{k})} \left( \frac{1}{e^{-\beta E(\vec{k})}+1}-\frac{1}{e^{\beta E(\vec{k})}+1} \right) \\
 & \Delta =  \frac{2V\alpha}{N} \sum_{k} \frac{ ( \cos k_x - \cos k_y)^2}{E(\vec{k})} \left( \frac{1}{e^{-\beta E(\vec{k})}+1}-\frac{1}{e^{\beta E(\vec{k})}+1} \right)
\end{align}
where $E(\vec{k})=\sqrt{\epsilon^2(\vec{k}) + \Delta^2(\vec{k}) + m^2U^2}$ with kinetic energy
$\epsilon(\vec{k})=-2t \left( \cos k_x + \cos k_y \right)$, and the gap equation $\Delta(\vec{k})=-2V\alpha \left( \cos k_x - \cos k_y \right) $.
We solve the fermion and the rotor mean-field equations self-consistently to obtain the eventual mean-field solution. In the calculation, we set $t=1$, $J=1$, $V/t=0.5$, and explore the $U/t-K/J$ ground state phase diagram as shown in Fig.~1(b) of the main text. 

\section{Determinantal quantum Monte Carlo simulation}
In this section, we present the details of the Determinantal Quantum Monte Carlo (DQMC) algorithm. Let's consider a new basis, $(\tilde{c}_{i,\uparrow}^{\dagger}, \tilde{c}_{i,\downarrow}^{\dagger}) \equiv  (c_{i,\uparrow}^{\dagger}, \epsilon_i c_{i,\downarrow})$, then the Hamiltonian becomes $H=H_t+H_V+H_U+H_{XY}$, with 
\begin{align}
& H_{t}= -t\sum_{\langle ij\rangle,\sigma}(\tilde{c}_{i,\sigma}^{\dagger}\tilde{c}_{j,\sigma}+\text{h.c.}) \\
& H_{V}= -V\sum_{\langle ij\rangle}(\tau_{i,j}e^{\i\theta_{i,j}}\epsilon_i(\tilde{c}_{i,\uparrow}^{\dagger}\tilde{c}_{j,\downarrow}-\tilde{c}_{j,\uparrow}^{\dagger}\tilde{c}_{i,\downarrow})+\text{h.c.}) \\
& H_{U}= -\frac{U}{2}\sum_{i}(\tilde{\rho}_{i,\uparrow}+\tilde{\rho}_{i,\downarrow}-1)^2 \\
& H_{XY}= K\sum_{\langle ij\rangle}n_{ij}^{2}-J\sum_{\langle ij,kl\rangle}\cos(\theta_{ij}-\theta_{kl}),
\end{align}
where $\tilde{\rho}_{i,\sigma} = \tilde{c}_{i,\sigma}^\dagger \tilde{c}_{i,\sigma}$. We consider a square lattice with system size $N=L\times L$.
To simulate this model by DQMC, we start with the partition function $\mathcal{Z}=\text{Tr}(e^{-\beta H})$, and perform the Trotter decomposition to divide the imaginary time evolution into $L_\tau$ slices $\beta=L_\tau \Delta_\tau$. Then we do further Trotter decomposition to separate the hopping part, the coupling part and the Hubbard interaction part.
To deal with the Hubbard interaction part, we introduce the following Hubbard-Stratonovich (HS) transformation~\cite{Assaad2005}:
\begin{equation}
e^{\Delta_\tau\frac{U}{2}(\tilde{\rho}_{i,\uparrow}+\tilde{\rho}_{i,\downarrow}-1)^{2}}=\frac{1}{4}\sum_{\{s_{i}\}}\gamma(s_{i})e^{\alpha\eta(s_{i})\left(\tilde{\rho}_{i,\uparrow}+\tilde{\rho}_{i,\downarrow}-1\right)}
\end{equation}
with $\alpha=\sqrt{\Delta_\tau\frac{U}{2}}$, $\gamma(\pm1)=1+\sqrt{6}/3$,
$\gamma(\pm2)=1-\sqrt{6}/3$, $\eta(\pm1)=\pm\sqrt{2(3-\sqrt{6})}$,
$\eta(\pm2)=\pm\sqrt{2(3+\sqrt{6})}$. Then we have the following fermion bilinear at time slice $\tau$,
\begin{align}
H_{\tau}= & \sum_i \alpha\eta(s_{i,\tau})\left(\tilde{\rho}_{i,\uparrow}+\tilde{\rho}_{i,\downarrow}-1\right) \nonumber \\
 & + \Delta_\tau V\sum_{\langle ij\rangle}\left(\tau_{i,j}\epsilon_i e^{\i\theta_{ij}(\tau)}\left(\tilde{c}_{i\uparrow}^{\dagger}\tilde{c}_{j\downarrow}-\tilde{c}_{i\downarrow}\tilde{c}_{j\uparrow}^{\dagger}\right)+\text{h.c.}\right) \nonumber \\
 & +\Delta_\tau t\sum_{\langle ij\rangle}\left(\tilde{c}_{i,\alpha}^{\dagger}\tilde{c}_{j,\alpha}+\text{h.c.}\right) \nonumber \\
\equiv & \tilde{\mathbf{c}}^{\dagger}\mathbf{K}_{U}\tilde{\mathbf{c}}+\tilde{\mathbf{c}}^{\dagger}\mathbf{V}_{\Delta}\tilde{\mathbf{c}}+\tilde{\mathbf{c}}^{\dagger}\mathbf{K}_{t}\tilde{\mathbf{c}}
\end{align}
where $\tilde{\mathbf{c}}=(\tilde{c}_{1,\uparrow},\tilde{c}_{1,\downarrow},\cdots,\tilde{c}_{N,\uparrow},\tilde{c}_{N,\downarrow})^T$, and $\mathbf{K}_U$, $\mathbf{V}_{\Delta}$ and $\mathbf{K}_t$ are the corresponding matrices of the three kinds of fermion bilinears.
Consider the antiunitary time reversal transformation: $\tilde{c}_{i\uparrow}\rightarrow\tilde{c}_{i\downarrow}$,
$\tilde{c}_{i\downarrow}\rightarrow-\tilde{c}_{i\uparrow}$, $\sqrt{-1}\rightarrow-\sqrt{-1}$.
It is easy to check that $\tilde{\mathbf{c}}^{\dagger}\mathbf{K}_{U}\tilde{\mathbf{c}}$, $\tilde{\mathbf{c}}^{\dagger}\mathbf{V}_{\Delta}\tilde{\mathbf{c}}$ and $\tilde{\mathbf{c}}^{\dagger}\mathbf{K}_{t}\tilde{\mathbf{c}}$
are all  invariant under this transformation. This proves that tracing out fermions in the partition function does not lead to a sign problem~\cite{wuzhang2005}. Next, we will show that  $H_{\XY}$  contribution to the partition function is also sign-problem-free,
thereby proving that the full Hamiltonian is sign-problem-free.

Now let us consider  $H_{\XY}$. It is convenient to work in a basis where $\theta_{ij}$ is diagonal, $\theta|\theta\rangle = \theta |\theta\rangle$. As $[n_{ij},e^{\pm \i \theta_{ij}}]=\pm e^{\pm \i \theta_{ij}} $, $n_{ij}$ behaves like $n_{ij}=-\i\frac{\partial}{\partial \theta_{i,j}}$. Denoting eigenbasis of $n_{ij}$ as $|n_{ij}\rangle$, we have $\langle \theta|n\rangle=e^{\i\theta n}$. This implies
\begin{equation}
\langle \theta' | e^{-\Delta_\tau K n^2} |\theta \rangle \sim e^{\frac{1}{2\Delta_\tau K}\cos(\theta-\theta')}
\end{equation}
We can now evaluate the partition function
\begin{align}
\mathcal{Z} & =\text{Tr}\left(e^{-\beta H}\right) \nonumber \\
 & =\text{Tr}_{F}\sum_{\{\theta\}}\langle\theta_{\tau=1}|e^{-\Delta\tau H}|\theta_{\tau=L_\tau}\rangle\langle\theta_{\tau=L_\tau}|e^{-\Delta\tau H}|\theta_{\tau=L_\tau-1}\rangle\cdots\langle\theta_{\tau=2}|e^{-\Delta\tau H}|\theta_{\tau=1}\rangle \nonumber \\
 & = \sum_{\{\theta,s\}}e^{\frac{1}{2\Delta\tau K}\sum_{\langle ij\rangle,\tau}\cos(\theta_{ij,\tau+1}-\theta_{ij,\tau}))+\Delta\tau J\sum_{\langle ij,kl\rangle,\tau}\cos(\theta_{ij,\tau}-\theta_{kl,\tau})}\left(\prod_{i,\tau}\gamma(s_{i,\tau})e^{-\alpha\eta(s_{i,\tau})}\right)\det\left(\mathbf{1}+\prod_{\tau}\mathbf{B}_{\tau}\right)
\end{align}
with $\mathbf{B}_{\tau}=e^{\mathbf{K}_{U}}e^{\mathbf{V}_{\Delta}}e^{\mathbf{K}_{t}}$
The first part of the last equation corresponds to an anisotropic XY model. We denote the configurations $\{\theta,s\}$ as $\mathcal{C}$, and write
their weight as $\omega_{\mathcal{C}}=\omega_{\mathcal{C}}^{I}\omega_{\mathcal{C}}^{II}$,
with
\begin{align}
 & \omega_{\mathcal{C}}^{I}=e^{\frac{1}{2\Delta\tau K}\sum_{\langle ij\rangle,\tau}\cos(\theta_{ij,\tau+1}-\theta_{ij,\tau}))+\Delta\tau J\sum_{\langle ij,kl\rangle,\tau}\cos(\theta_{ij,\tau}-\theta_{kl,\tau})}\left(\prod_{i,\tau}\gamma(s_{i,\tau})e^{-\alpha\eta(s_{i,\tau})}\right) \\
& \omega_{\mathcal{C}}^{II}=\det\left(\mathbf{1}+\prod_{\tau}\mathbf{B}_{\tau}\right)
\end{align}
It is obvious that $\omega_\mathcal{C}^I$ is sign-problem-free, while  $\omega_{\mathcal{C}}^{II}$  was already shown to be sign-problem-free using antiunitary time reversal symmetry. Thus, as promised, the model we simulate is sign problem free.
\paragraph{Monte Carlo updates:} We update the auxiliary fields locally in DQMC. There are two kinds of auxiliary fields to be updated. For $s_{i,\tau}$, as it appears only along the diagonal of $\mathbf{K}_U$, a local change can be made as:
\begin{equation}
e^{\mathbf{K}_U'} = (\mathbf{1}+\boldsymbol{\Delta})e^{\mathbf{K}_U}
\end{equation}
where $\boldsymbol{\Delta}$ has only two nonzero elements $\boldsymbol{\Delta}_{i\uparrow,i\uparrow}$ and $\boldsymbol{\Delta}_{i\downarrow,i\downarrow}$. Let's define $\mathbf{B}(\tau',\tau)\equiv \mathbf{B}_{\tau'}\cdots \mathbf{B}_{\tau}$, and note that the equal-time Green's function $\mathbf{G}(\tau,\tau)=(\mathbf{1}+\mathbf{B}(\tau,0)\mathbf{B}(\beta,\tau))^{-1}$. With a local update, we have 
\begin{equation}
\frac{\omega_{\mathcal{C'}}^{II}}{\omega_{\mathcal{C}}^{II}} = \det [\mathbf{1}+\boldsymbol{\Delta}(\mathbf{1}-\mathbf{G}(\tau,\tau))]
\end{equation}
and the ratio $\frac{\omega_{\mathcal{C'}}^{I}}{\omega_{\mathcal{C}}^{I}}$ is straightforward to evaluate.
If the local update is accepted, we update the equal-time Green's function as
\begin{equation}
\mathbf{G}'(\tau,\tau) = \mathbf{G(\tau,\tau)}[\mathbf{1}+\boldsymbol{\Delta}(\mathbf{1}-\mathbf{G})]
\end{equation}
Due to the sparse nature of $\boldsymbol{\Delta}$, the fast update algorithm using Sherman-Morrison-Woodbury formula applies. The update for the $\theta$ fields are very similar, the only difference is that $\boldsymbol{\Delta}$ is different.
\paragraph{Measurements in Monte Carlo:} 
Given equal-time Green's functions, one can use Wick's theorem to measure various static quantities, such as the energy and correlation functions of spins, d-wave pairing order parameters, and so on. In addition to static calculations, we also calculate dynamic properties, using imaginary time-displaced Green's function, again using Wick's theorem. In particular, we measure the time-displaced single particle Green's functions and the time-displaced spin correlations. Further technical details about the updates and measurements in DQMC can be found in Refs.~\cite{BSS,AssaadEvertz2008}.

\section{Charge Stiffness at the XY transition}
As discussed in the main text, the transition from dSC$_\text{g}$+AFM to AFM is a 3D XY transition, which predicts universal behavior of charge stiffness with the tuning parameter. In the following, we derive the formula to measure the charge stiffness following the method described in Ref.~\cite{Van1996}. 

By definition, the stiffness measures the free-energy increment associated with twisting the direction of the order parameter. For the transition from dSC$_\text{g}$+AFM to AFM, the order parameter can be defined as $\langle e^{i\theta_{ij}} \rangle$. We consider a twist $\theta_{ij} \rightarrow \theta_{ij}-\vec{q}\cdot \vec{r}_{ij}$, and the stiffness is calculated as the second derivative of the free-energy with respect to this twist:
\begin{equation}
\bar{\bar{\rho}}_{\text{s}} =\frac{1}{N}\left.\frac{\partial^{2}F(\phi)}{\partial\vec{q}\partial\vec{q}}\right|_{\vec{q}=0}
\end{equation}
More explicitly, we
expand the relevant part of the Hamiltonian in powers of $i\vec{q}$, so that we have
\begin{equation}
H_{\XY}'=H_{\XY}+\i\vec{q}\cdot\vec{H}_{\XY,1}+\frac{1}{2}\i\vec{q}\cdot\bar{\bar{H}}_{\XY,2}\cdot \i\vec{q}
\end{equation}
where $\vec{H}_{\XY,1}=\i J\sum_{\langle ij,kl\rangle}\vec{r}_{ij,kl}\sin(\theta_{ij}-\theta_{kl})$ can be thought of as a current operator and
$\bar{\bar{H}}_{\XY,2}=-J\sum_{\langle ij,kl\rangle}\vec{r}_{ij,kl}\vec{r}_{ij,kl}\cos(\theta_{ij}-\theta_{kl})$. The free energy increment due to $\vec{q}$ can be calculated using
perturbation theory
\begin{equation}
\delta F=\frac{1}{2}\i\vec{q}\cdot\left\{ \langle\bar{\bar{H}}_{\XY,2}\rangle-\int_0^\beta\left\langle \vec{H}_{\XY,1}(\tau) \vec{H}_{\XY,1}(0)\right\rangle\right\} \cdot \i\vec{q}
\end{equation}
Where we have used that $\langle\vec{H}_{XY,1}\rangle=0$. Thus we obtain 
\begin{equation}
\bar{\bar{\rho}}_{s}=-\frac{1}{N} \left\{ \langle\bar{\bar{H}}_{\XY,2}\rangle-\int_0^\beta\left\langle \vec{H}_{\XY,1}(\tau) \vec{H}_{\XY,1}(0)\right\rangle\right\} 
\end{equation}
Due to the $\pi/2$ rotation symmetry of the square lattice, $\bar{\bar{\rho}}_\text{s}=\rho_{s}\bar{\bar{I}}$, and we can consider the twist along either direction. Let us take $x$-direction as an example, and we obtain
\begin{equation}
\rho_{s}=-\frac{1}{N} \left\{ \langle H^x_{\XY,2}\rangle-\int_0^\beta\left\langle H^x_{\XY,1}(\tau) H^x_{\XY,1}(0)\right\rangle\right\} 
\end{equation}
where $H^x_{\XY,1}=\i J\sum_{\langle ij,kl\rangle}x_{ij,kl}\sin(\theta_{ij}-\theta_{kl})$,
$H^x_{\XY,2}=-J\sum_{\langle ij,kl\rangle}\frac{1}{4}\cos(\theta_{ij}-\theta_{kl})$.

\subsection{Stiffness of the pure rotor model}
\begin{figure}[h]
  \centering
	\includegraphics[width=0.6\hsize]{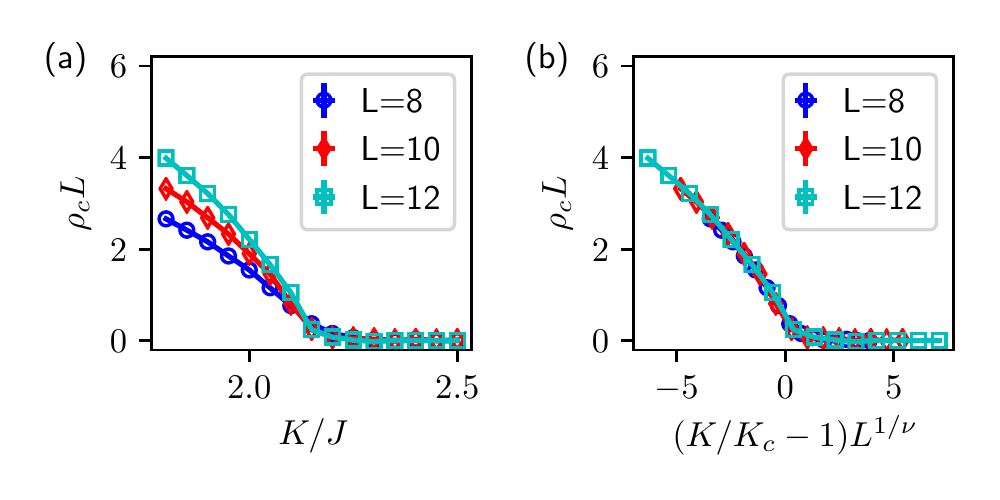}
	\caption{Charge stiffness of the pure (2+1)D rotor model. $K_{c} \approx 2.13$, $\nu\approx0.67$.}
	\label{fig:stfxy}
\end{figure}

To test above formula within our simulations, we first consider the pure rotor model described by the Hamiltonian $H_{XY}$. As is well-known, and we already showed above explicitly using Trotter decomposition, the (2+1)D pure rotor model can be
mapped to 3D XY model so that it has a 3D XY transition. Therefore we expect the associated universal critical behavior of the stiffness with respect to the tuning parameter $K/J$. As shown in Fig.~\ref{fig:stfxy}, our simulation of this pure rotor model indeed finds that  the data for stiffness can be collapsed with the 3D XY exponents.

\subsection{Stiffness of interacting model and finite temperature transition}
\begin{figure}[h]
  \centering
	\includegraphics[width=0.9\hsize]{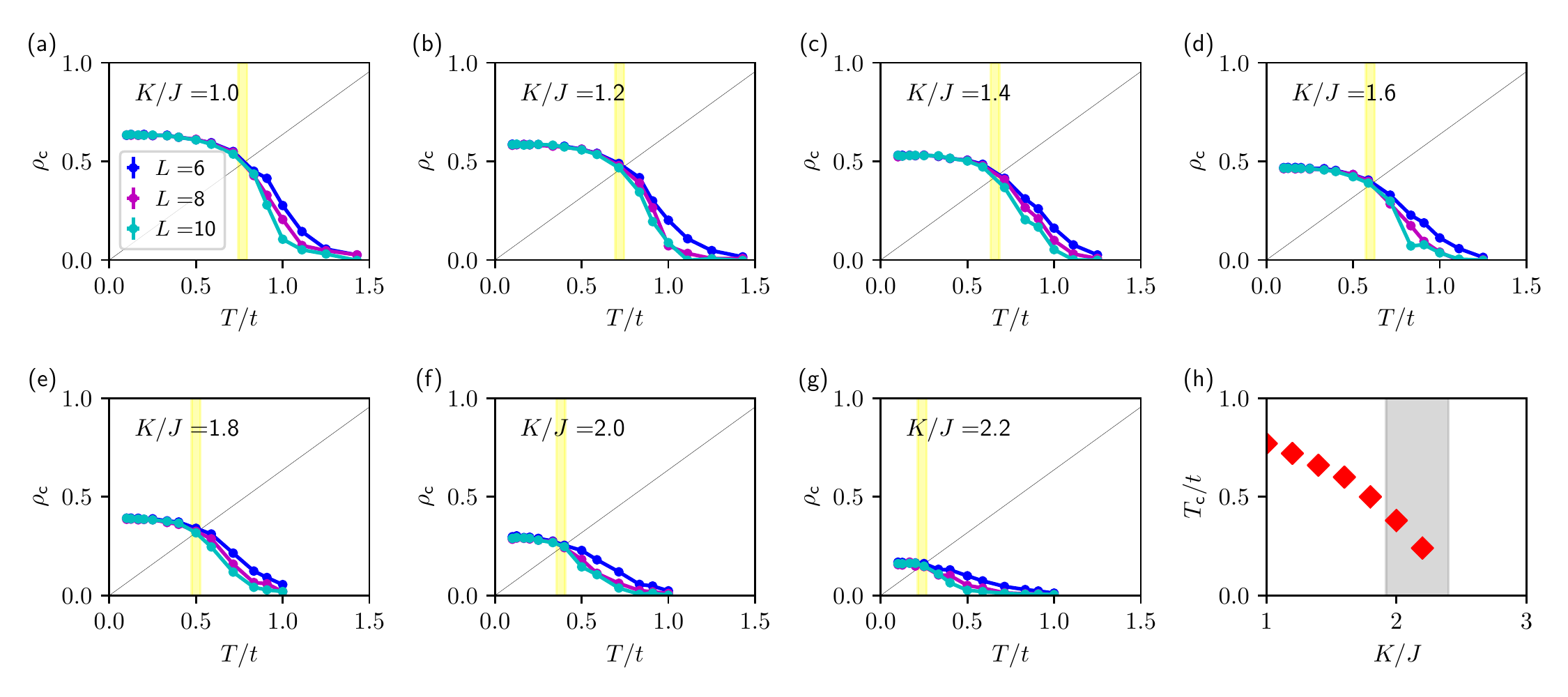}
	\caption{Using charge stiffness to determine BKT phase transition boundary. (a)-(g) Charge stiffness for $K/J=1.0,\ 1.2,\ 1.4,\ 1.6,\ 1.8,\ 2.0,\ 2.2$. (h) BKT phase transition boundary.}
	\label{fig:stif}
\end{figure}

The stiffness of our full, interacting model with $\beta t=2L$ is shown in Figs.~4(a) and (b), which is still well collapsed with 3D XY exponents, indicating that the transition is continuous and belongs to the 3D XY universality. The charge stiffness is also a good quantity to characterize finite temperature Berezinskii-Kosterlitz-Thouless (BKT) transition phase boundary. Based on the free energy argument of a single vortex, the BKT transition is determined by $\rho_\text{c}=\frac{2T_\text{c}}{\pi}$, see Figs.~\ref{fig:stif}(a)-(g). The BKT phase boundary is plotted in Fig.~\ref{fig:stif}(h).

\section{Determining zero temperature phase diagram}
\subsection{dSC order parameters}
\begin{figure}[h]
  \centering
	\includegraphics[width=0.5\hsize]{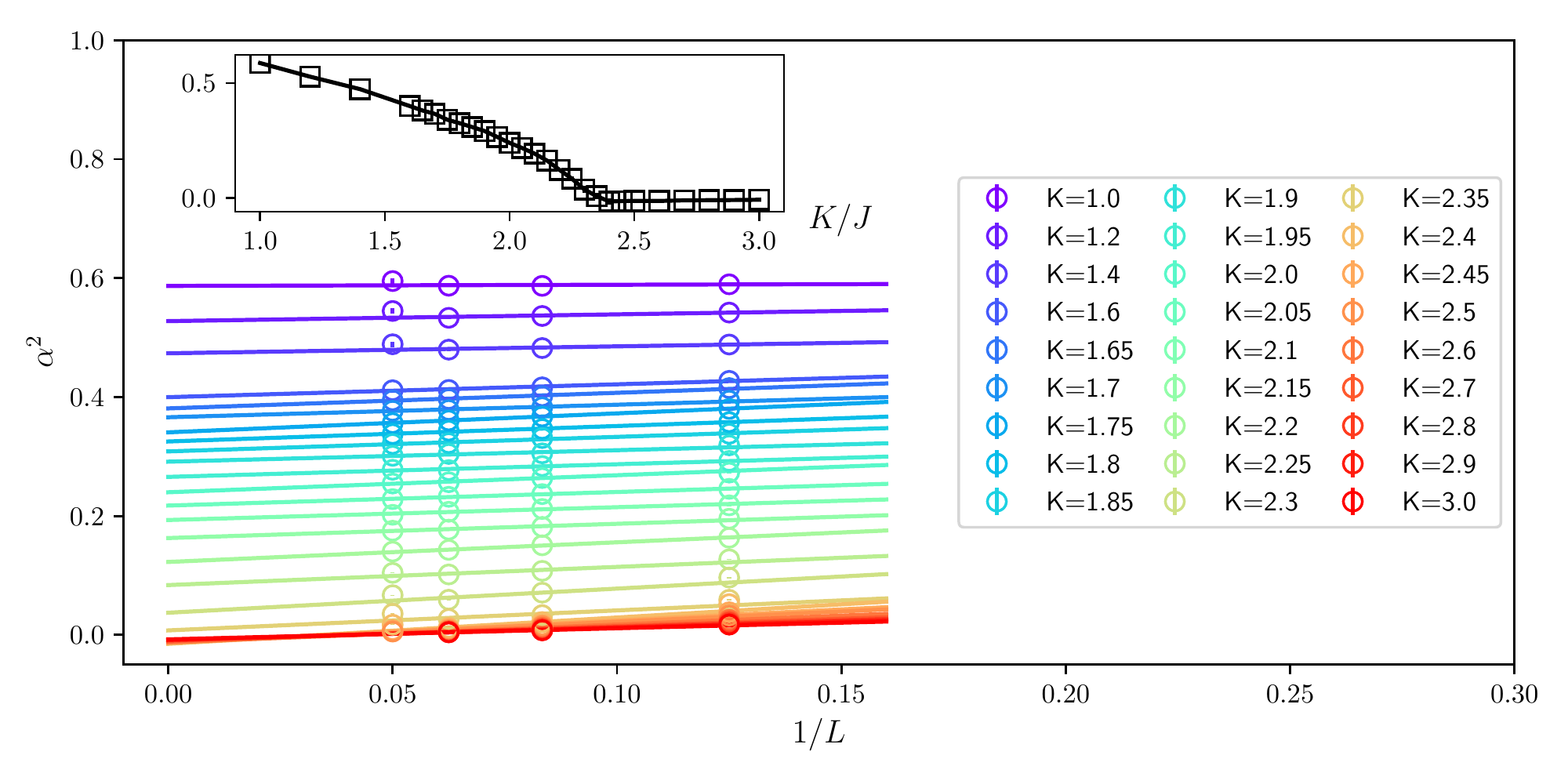}
	\caption{Finite size scaling of $\alpha^2$.}
	\label{fig:a2}
\end{figure}

The dSC order parameter $\alpha$ defined as $\alpha = \langle \hat{\alpha}_{ij} \rangle$ can be measured through the static correlation function
\begin{equation}
\alpha^2 = \frac{1}{4L^4} \sum_{\langle ij \rangle, \langle k l \rangle} \langle e^{\i \theta_{ij}} e^{\i \theta_{kl}} \rangle.
\end{equation}
We perform a finite size scaling of $\alpha^2$ to estimate its value in the thermodynamic limit, as shown in Fig.~\ref{fig:a2}. We find that the dSC order parameter vanishes for $K/J \geq 2.40(5)$.

\subsection{AFM order parameters}
\begin{figure}[h]
  \centering
	\includegraphics[width=0.5\hsize]{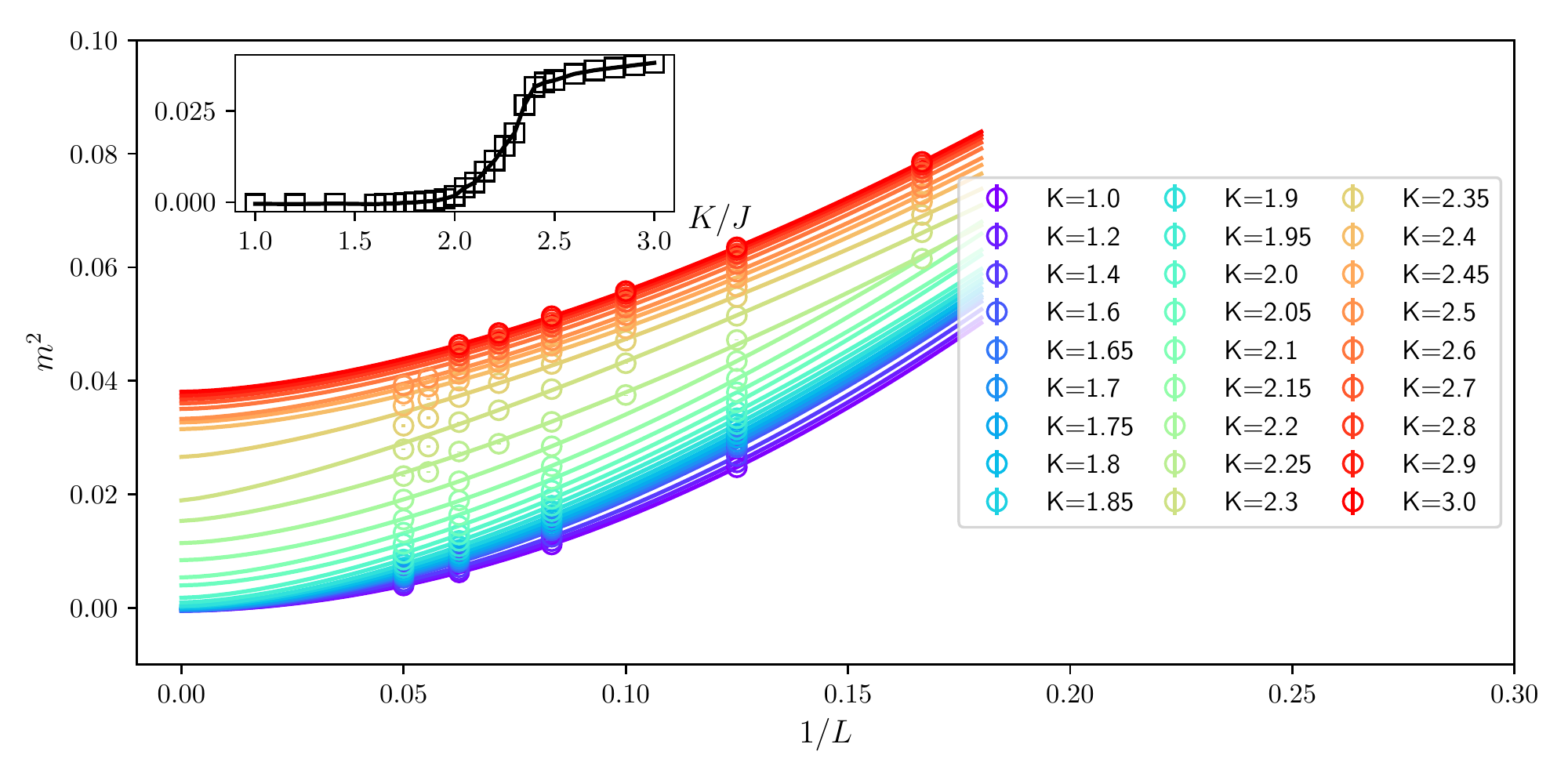}
	\caption{Finite size scaling of $m^2$.}
	\label{fig:m2}
\end{figure}

The AFM order parameter defined as $\vec{m}=\langle (-)^{i} \vec{S}_i \rangle$ can be measured through static correlation function
\begin{equation}
m^2 = \frac{1}{L^4} \sum_{i,j} (-)^{i+j} \left\langle\vec{S}_i \cdot \vec{S}_j \right\rangle.
\end{equation}
Again, we perform a finite size scaling of $m^2$ to estimate its value in the thermodynamic limit, as shown in Fig.~\ref{fig:m2}. The AFM order parameter vanishes for $K/J \leq 1.92(5)$.

\subsection{Single-particle gap}
\begin{figure}[h]
  \centering
	\includegraphics[width=0.5\hsize]{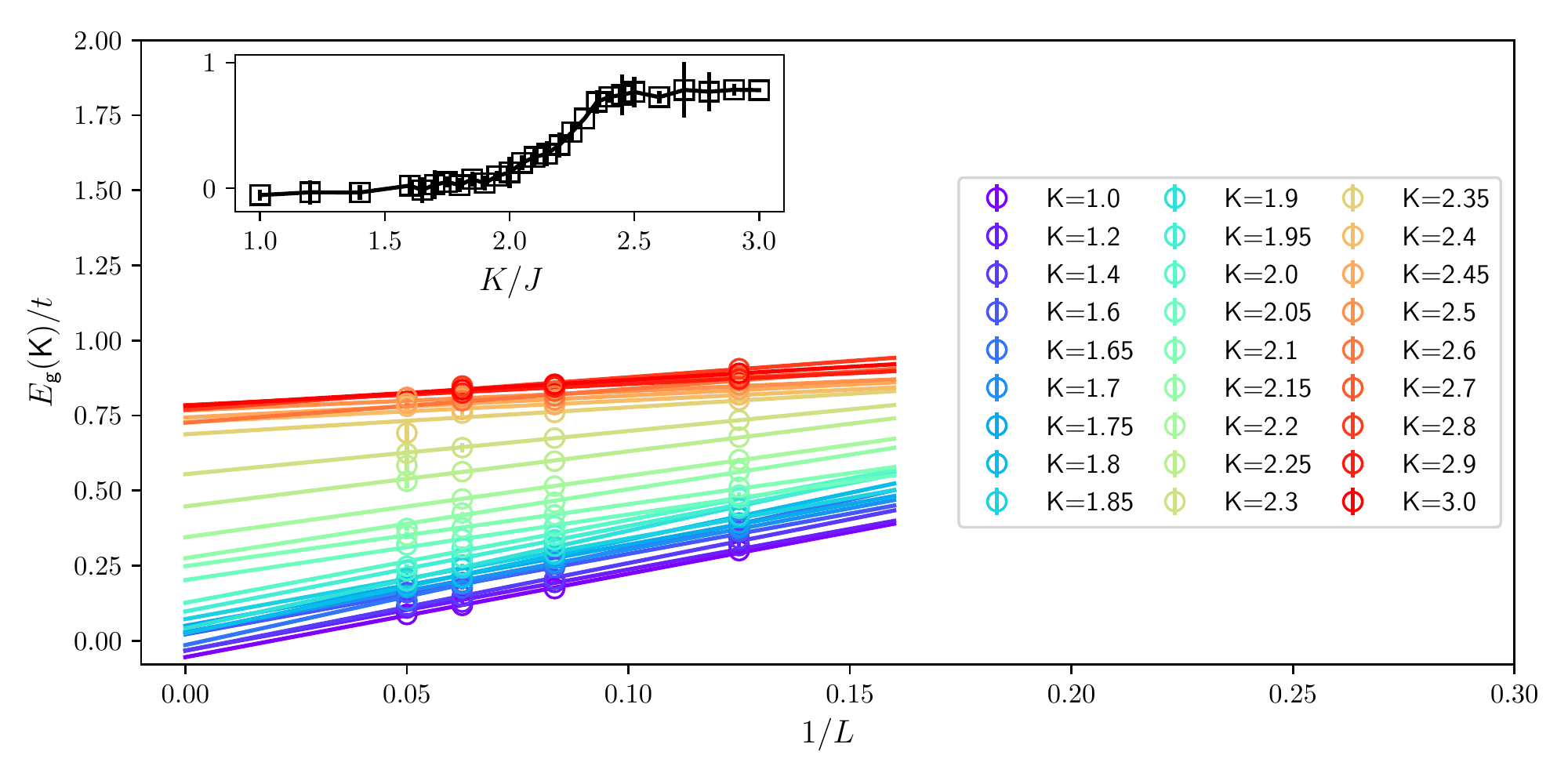}
	\caption{Finite size scaling of the single-particle gap at the K point.}
	\label{fig:Kgap}
\end{figure}

The single-particle gap $\Eg(\vec{k})$ can be measured by extracting the slope of the decay of the time-displaced single particle Green's functions, based on the equation
\begin{equation}
\langle c(\vec{k},\tau)c^\dagger(\vec{k},0) \rangle \sim e^{-\Eg(\vec{k})\tau},
\end{equation}
where a sum over spin index is implicit. After we obtain the single-particle gap for each finite size $L$, we perform a $1/L$ extrapolation to obtain its value in the thermodynamic limit. In Fig.~\ref{fig:Kgap} we show the single particle gap at the K point. The single-particle gap opens around $K/J=1.92(5)$ where the AFM order parameter starts to rise. Therefore, between $K/J=1.92(5)$ and $K/J=2.40(5)$, we have a coexistence of gapped dSC and AFM, which is denoted as the dSC$_\text{g}$+AFM phase in the main text.

\begin{figure}[h]
  \centering
	\includegraphics[width=0.5\hsize]{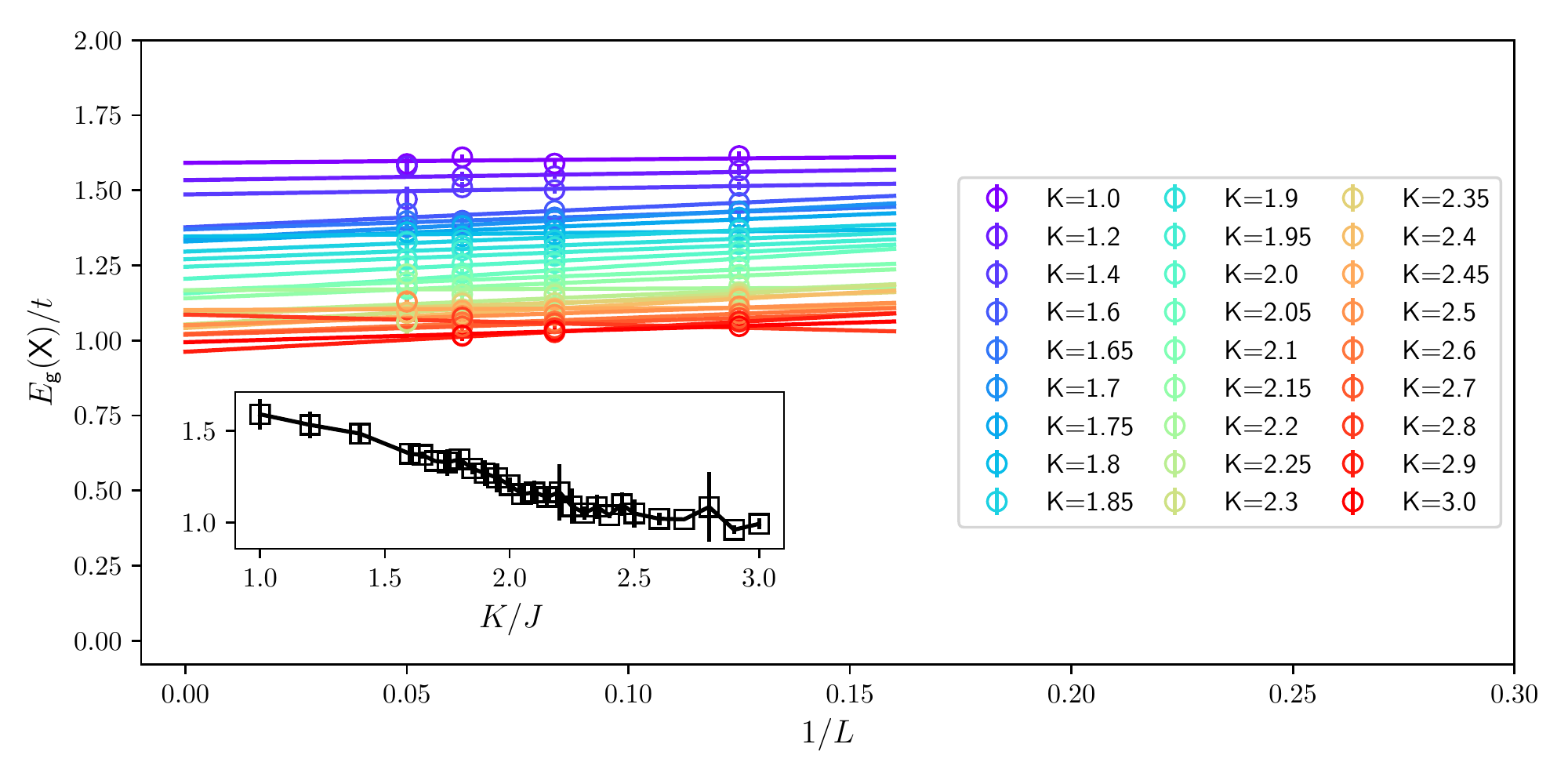}
	\caption{Finite size scaling of single-particle gap at the X point.}
	\label{fig:Xgap}
\end{figure}

On the other hand, we still need to confirm that the dSC phase with $K/J$ less than $1.92(5)$ is nodal, and that a full Fermi surface is absent. We explore the single-particle gap along different momentum paths, and confirm that only the four K points are gapless, as shown in Fig.~3(c). We also perform a more rigorous finite size scaling of the single-particle gap at X point, the `anti-nodal' point, to confirm that this point is indeed gapped in the nodal dSC phase, as shown in Fig.~\ref{fig:Xgap}. 

\begin{figure}[h]
  \centering
	\includegraphics[width=0.5\hsize]{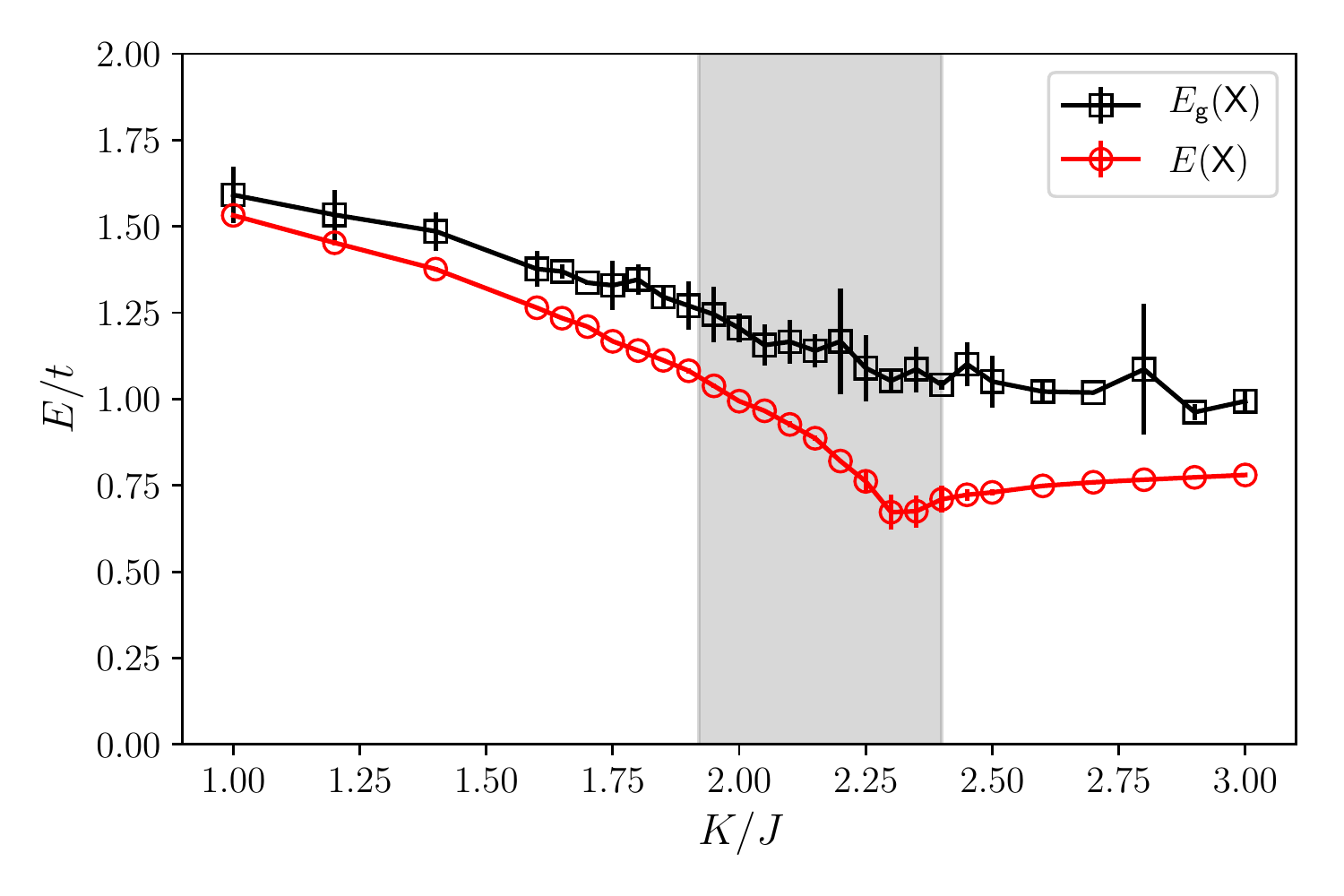}
	\caption{Comparison of the  the mean-field predicted anti-nodal gap with the estimation from time-displaced Green's functions.}
	\label{fig:Xgapcompare}
\end{figure}

Furthermore, based on the mean-field theory, the anti-nodal gap is predicted to be $E(\text{X})=\sqrt{16V^2\alpha^2+m^2 U^2}$. As we already extracted the value of both $\alpha^2$ and $m^2$ using DQMC, we can compare the mean-field predicted anti-nodal gap with the estimation from time-displaced Green's functions. As shown in Fig.~\ref{fig:Xgapcompare}, the mean-field estimate and the actual gap overall track each other rather well. The deviation is larger close to the XY transition, presumably because of larger fluctuations.

\section{Spin gap}
Similar to the single-particle gap, the spin-gap $\Es(\vec{k})$ can be measured by extracting the slope of the decay of the time-displaced spin correlation functions using
\begin{equation}
\langle \vec{S}(\vec{k},\tau)\cdot \vec{S}(-\vec{k},0) \rangle \sim e^{-\Es(\vec{k})\tau}.
\end{equation}

\section{Extracting Fermi velocity, pairing velocity and spin-wave velocity}
\begin{figure}[h]
  \centering
	\includegraphics[width=0.5\hsize]{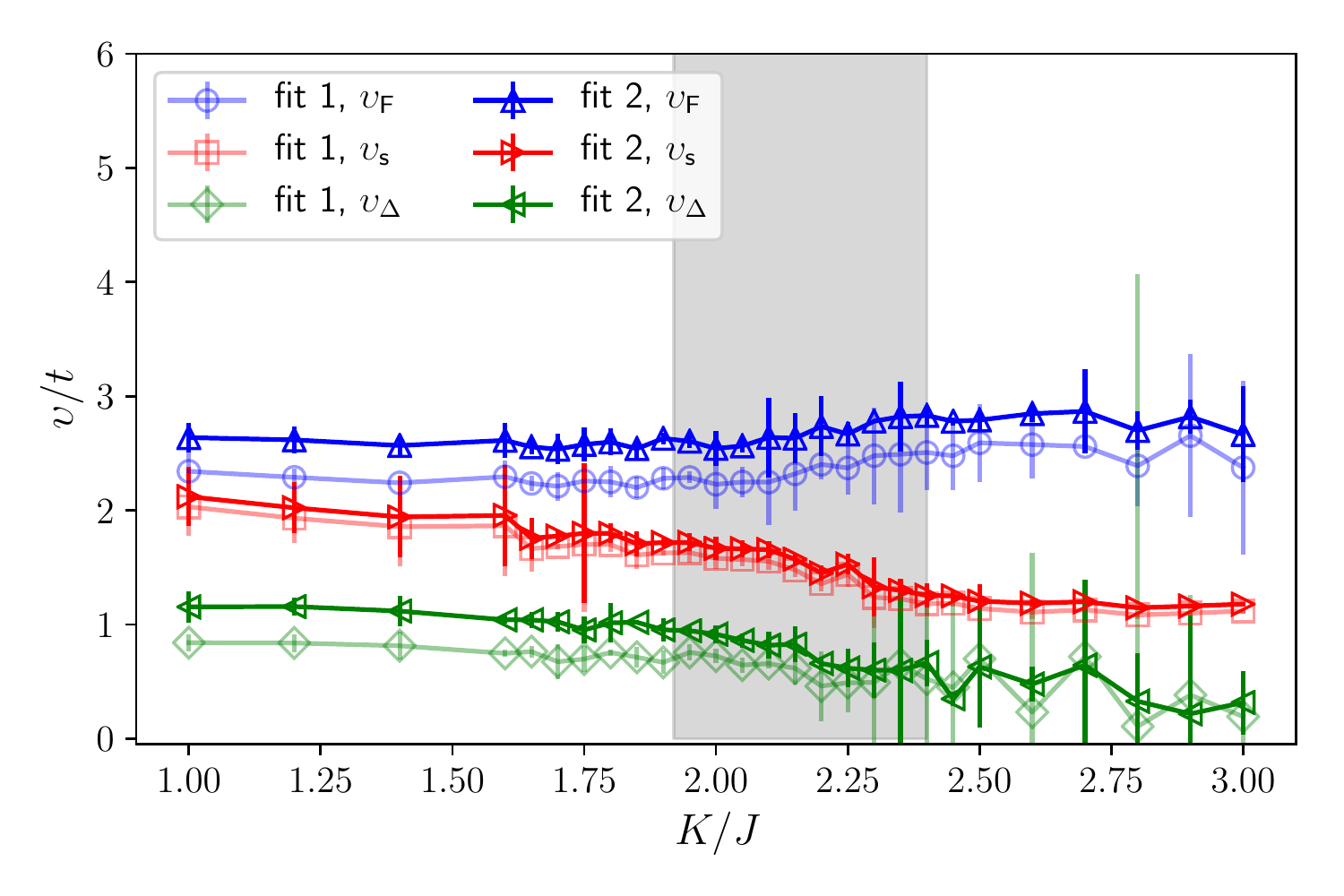}
	\caption{The three velocities $\upsilon_F, \upsilon_s, \upsilon_{\Delta}$ obtained using the two schemes discussed in the main text (Fig.~3(e)).}
	\label{fig:vel}
\end{figure}
\begin{figure}[h]
  \centering
	\includegraphics[width=0.4\hsize]{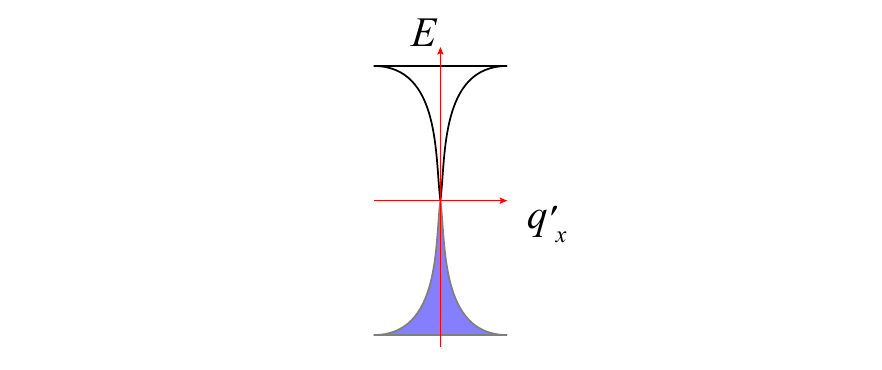}
	\caption{The curvature of the Dirac dispersion can lead to finite size effects in extracting the velocities, as discussed in the accompanying text.}
	\label{fig:vel_fs}
\end{figure}

Based on the estimation of the single-particle gap $\Eg(\vec{k})$ and the spin-gap $\Es(\vec{k})$, we can extract the Fermi velocity $\upsilon_\text{F}$, the pairing velocity $\upsilon_\Delta$ and the spin-wave velocity $\upsilon_\text{s}$. Denote $\vec{\dk}_1$ as the smallest momentum along the K-M direction, and $\vec{\dk}_2$ the smallest momentum along the K-Y direction (see Fig.~3(a) for the definition of various points in the Brillouin zone). Based on the low energy theory, we have $\Eg(\vec{q}')=\sqrt{\upsilon_F^2q_x'^2+\upsilon_\Delta^2q_y'^2+E_g^2(\text{K})}$ and $\Es(\vec{q}')=\sqrt{\upsilon_s^2 q'^2 + \Es^2(\text{M})}$. Therefore we can use following formulas to extract velocities:
\begin{align}
\vF & = \sqrt{\frac{\Eg^2(\text{K}+\vec{\dk}_1)-\Eg^2(\text{K})}{|\vec{\dk}_1|^2}} \label{eq:vf}\\
\vD & = \sqrt{\frac{\Eg^2(\text{K}+\vec{\dk}_2)-\Eg^2(\text{K})}{|\vec{\dk}_2|^2}} \label{eq:vd}\\
\vs & = \sqrt{\frac{\Es^2(\text{M}+\vec{\dk}_1)-\Es^2(\text{M})}{|\vec{\dk}_1|^2}} \label{eq:vs}
\end{align}
with $\vec{\dk}_1=(\delta q,\delta q)$, $\vec{\dk}_2=(-\delta q,\delta q)$ and $\delta q=\frac{2\pi}{L}$. We considered two schemes to perform the finite size scaling. In scheme 1, we fix $\delta q = \frac{2\pi}{L_\text{max}}$, and for system sizes less than $L_\text{max}=20$, $\Eg(\text{K}+\vec{\dk}_1)$, $\Eg(\text{K}+\vec{\dk}_2)$ and $\Es(\Gamma+\vec{\dk}_1)$ are obtained by interpolation. After we obtain the gap functions for each system size, we perform an $1/L$ extrapolation of the gap, and finally use the above formulas to obtain the velocities.
In scheme 2, we first calculate the velocities based on the above formulas for different system sizes $L$ (that is $\delta q=\frac{2\pi}{L}$ is not fixed), and then perform a $1/L$ extrapolation of the velocities. The velocities we get from both schemes are shown in Fig.~\ref{fig:vel}, which is a zoomed out version of Fig.~3(e). In principle, in the large $L_\text{max}$ limit, both schemes should yield same value for the velocities. However, in finite size calculations, as we can see that the scheme 1 comparatively yields smaller values. This may be understood as follows: the Dirac cones have a rather  large curvature near the nodes as sketched in Fig.~\ref{fig:vel_fs}. Consider  expanding the energy $E$ to the second order in $q_x'$ near the node, $E=\upsilon_F q_x' + \frac{1}{2}c q_x'^2$ with curvature $c<0$ and we have assumed $q_x' \geq 0$ for concreteness. At this order, for scheme 1 we find $\upsilon_F^{\text{scheme 1}}=\upsilon_F + \frac{c \pi}{L_{\text{max}}} < \upsilon_F$.
In contrast, for scheme 2 we find $\upsilon_F^{\text{scheme 2}}=\upsilon_F + \frac{c \pi}{L}$ for each system size $L$, and it extrapolates to $\upsilon_F$ when $1/L$ is extrapolated to zero.

\section{Analytical continuation to get single particle and spin spectrum}
We use stochastic analytical continuation method~\cite{Sandvik1998} to extract the real frequency spectrum of both the single-particle and the spin. The analytic continuation is based on solving an inverse problem: given the imaginary time Green's functions $G(\tau)$, the task is to obtain the real frequency spectrum $\rho(\omega)$. These quantities are related by
\begin{equation}
G(\tau) = \int \frac{d\omega}{2\pi} \frac{\rho(\omega)e^{-\omega \tau}}{e^{-\beta\omega}\pm 1}
\end{equation}
where $\pm$ is for the fermionic and the bosonic case, respectively. In Fig.~3(a), we show the single-particle spectrum integrated over a small energy window near the Fermi level to find the location of the Dirac nodes inside the nodal dSC phase.

\section{Autocorrelation time and critical slowing down}

As in any standard DQMC simulation, the Monte Carlo moves are local and therefore, in principal the simulations suffer from the critical slowing down close to the two transitions. At the same time, since we are limited to sizes of the order of $20 \times 20$ (due to $\beta N^3$ scaling of DQMC), the critical slowing down is not a serious impedance. In particular, the autocorrelation time for the transition from dSC to coexistence phase is of the order of a few hundred sweeps for our largest system sizes (e.g. at $L = 20$, it is about 300 sweeps), while for the transition from the coexistence phase to the AFM transition, it is about one thousand sweeps at $L = 20$.

\end{document}